\documentclass[12pt]{article}

\usepackage{amssymb}
\usepackage{amsmath}
\usepackage{amsfonts}
\usepackage{geometry}
\usepackage{color}
\usepackage{comment}
\usepackage{footmisc}
\usepackage{subcaption}
\usepackage{hyperref}
\usepackage[inline]{enumitem}
\usepackage[toc,page]{appendix} % For adding appendices
\usepackage{tocloft} % optional, only if you want TOC formatting tweaks
\usepackage{graphicx}
\usepackage{mathptmx}
\usepackage[singlespacing]{setspace}
\usepackage{fancyhdr}
\usepackage{natbib}
\usepackage{ulem}
\usepackage{xcolor}
\usepackage{array}
\usepackage[most]{tcolorbox}
\usepackage{booktabs}
\usepackage{tabularx}
\usepackage{multirow}
\usepackage{makecell}
\usepackage{mathtools}

\usepackage[linesnumbered,ruled,vlined]{algorithm2e}
\setcitestyle{authoryear,open={(},close={)}}
\graphicspath{{images/}{images/si/}}

\newtheorem{theorem}{Theorem}

\newtheorem{proposition}{Proposition}
\newenvironment{proof}[1][Proof]{\noindent\textbf{#1.} }{\ \rule{0.5em}{0.5em}}

\newcolumntype{L}[1]{>{\raggedright\let\newline\\arraybackslash\hspace{0pt}}m{#1}}
\newcolumntype{C}[1]{>{\centering\let\newline\\arraybackslash\hspace{0pt}}m{#1}}
\newcolumntype{R}[1]{>{\raggedleft\let\newline\\arraybackslash\hspace{0pt}}m{#1}}

\newcommand{\finalTotalUserN}{73,136}
\newcommand{\noActivityBadStatusUserN}{7,018}

\newcommand{\noActivityUserN}{2,369}

\newcommand{\badStatusUserN}{4,649}

\newtcolorbox{outlinebox}{
  colback=gray!10, colframe=black,
  title=OUTLINE,
  boxrule=0.8pt,
  arc=4pt,
  left=6pt, right=6pt, top=6pt, bottom=6pt,
}

\title{
\vspace{-0.2em}
Celebrity messages reduce online hate and limit its spread\thanks{
We have benefited from seminar and conference feedback at the Oxford Internet Institute's Future of Social Media Research workshop,  IC2S2 23, Meta's CSS Seminar, and the World Bank DECDI seminar. This study was conducted as part of the World Bank's Development Impact research program. 
\textbf{Author contributions:}
E.J., and S.F. led, directed, and oversaw the project. H.L. managed the ad campaign. D.B. collected the data with contributions from H.L. M.T. computed hate scores. B.K. and E.J. prepared the data with contributions from M.T., D.B. and H.L. B.K. conducted descriptive analysis with contributions from M.T. E.J. conducted all other data analysis with contributions from H.L. E.J. and S.F. led the writing of the manuscript with contributions from B.K., M.T., H.L., and V.O. All authors contributed to designing the research.
}
}

\author{
Eaman Jahani\thanks{To whom correspondence may be addressed. Email: \url{eaman@umd.edu}; \url{sfraiberger@worldbank.org}} \\ University of Maryland and MIT IDE
\and Blas Kolic \\ Universidad Carlos III de Madrid
\and Manuel Tonneau \\ University of Oxford
\and Hause Lin \\ MIT
\and Daniel Barkoczi \\ University of Southern Denmark
\and Edwin Ikhuoria \\ Middlesex University \\
\and Victor Orozco \\ World Bank
\and Samuel P. Fraiberger \\ World Bank and NYU \footnotemark[2]
}

\date{}

\begin{document}

\maketitle
\vspace{-0.9em}
\begin{abstract}
\noindent Online hate spreads rapidly, yet little is known about whether preventive and scalable strategies can curb it. We conducted the largest randomized controlled trial of hate speech prevention to date: a 20-week messaging campaign on X in Nigeria targeting ethnic hate. \finalTotalUserN{} users who had previously engaged with hate speech were randomly assigned to receive prosocial video messages from Nigerian celebrities. The campaign reduced hate content by 2.5\% to 5.5\% during treatment, with about 75\% of the reduction persisting over the following four months. Reaching a larger share of a user’s audience reduced amplification of that user’s hate posts among both treated and untreated users, cutting hate reposts by over 50\% for the most exposed accounts. Scalable messaging can limit online hate without removing content.

\vspace{0in}

\bigskip
\end{abstract}

\setcounter{page}{0}
\thispagestyle{empty}
\pagebreak \newpage

\doublespacing

\section*{Introduction} \label{sec:introduction}

Social media platforms have reshaped how information circulates, enabling rapid coordination during social movements, protests, and civic engagement \citep{bond2012facebook,tufekci2017twitter,enikolopov2020social}. Yet the same attention-driven systems also amplify hateful content \citep{vosoughi2018spread,bail2018exposure,finkel2020,guess2020exposure,milli2025amplification}. Global surveys indicate that about two-thirds of internet users report encountering hate speech online, particularly on social media \citep{unesco_ipsos_2023_disinformation_hate_speech}. Prior evidence links online hate to offline harms, including ethnic persecution in Myanmar, attacks on refugees in Germany, and hate crimes against migrants and LGBT communities in Spain \citep{muller2021fanning,whitten2020myanmar,arcila2024online}.

In response, platforms have largely relied on content moderation, removing or restricting rule-violating posts to limit exposure \citep{gillespie2018custodians}. Although moderation can reduce hate and improve compliance with platform rules \citep{chandrasekharan2017you,kumarswamy2023impact,yildirim2023short,horta2023automated}, it faces important limitations. Harmful material can spread more quickly than it can be reviewed, enforcement is uneven across users and languages, and human review is costly to scale \citep{vosoughi2018spread,pfeffer2023half,giansiracusa2021facebook,goldstein2023understanding,tonneau-etal-2025-hateday,tonneau2025language}. Automated systems can misclassify content, sometimes missing less overt material or removing non-harmful posts, which raises free-expression concerns \citep{allen2024quantifying,lee2024people,meta2025more_speech}. These limits highlight the need for scalable approaches that reduce online hate without removing content.

One tested non-censoring approach is direct counterspeech, which involves publicly replying to hateful posts to discourage hostility. Prior work shows that such replies can reduce hateful expression among treated users \citep{munger2017tweetment,siegel2020no2sectarianism,hangartner2021empathy,bar2024generative,gennaro2025counterspeech}. Yet direct counterspeech is inherently reactive, requiring hate content to appear and be detected before any response, often after it has already spread. Scaling such interventions is also difficult because they rely on real-time detection across large volumes of content, a capability largely limited to platforms.

A complementary approach is preventive counterspeech, which aims to act before hateful content spreads by reaching users who are at risk of producing or sharing it. Targeted messaging delivered through ads offers a proactive and scalable channel and has shown promise in reducing misinformation sharing and increasing vaccine uptake \citep{lin2024reducing,larsen2022using}. Effects may be amplified when messages feature celebrities, who can increase attention, engagement, and credibility \citep{alatas2023celebrities}. However, no large-scale study has tested whether preventive messaging can reduce online hate or generate broader network effects.

We report the largest field experiment to date testing a preventive strategy to reduce online hate. In a preregistered randomized controlled trial on X (formerly Twitter) in Nigeria, we delivered a 42-second prosocial video featuring Nigerian celebrities to \finalTotalUserN{} users who had previously engaged with hate content. Nigeria provides a salient test case given widespread social media use and persistent inter-ethnic tensions \citep{ezeibe2021hate}. We evaluate the effects on hate production and sharing, their persistence after messaging ends, and whether reaching a large share of a user’s audience generates indirect effects across the network. The results show that targeted celebrity messaging can reduce online hate and curb its spread without removing content.

\section*{Results} \label{sec:results}

\begin{figure}[t!]
    \centering
    \includegraphics[width=1\linewidth]{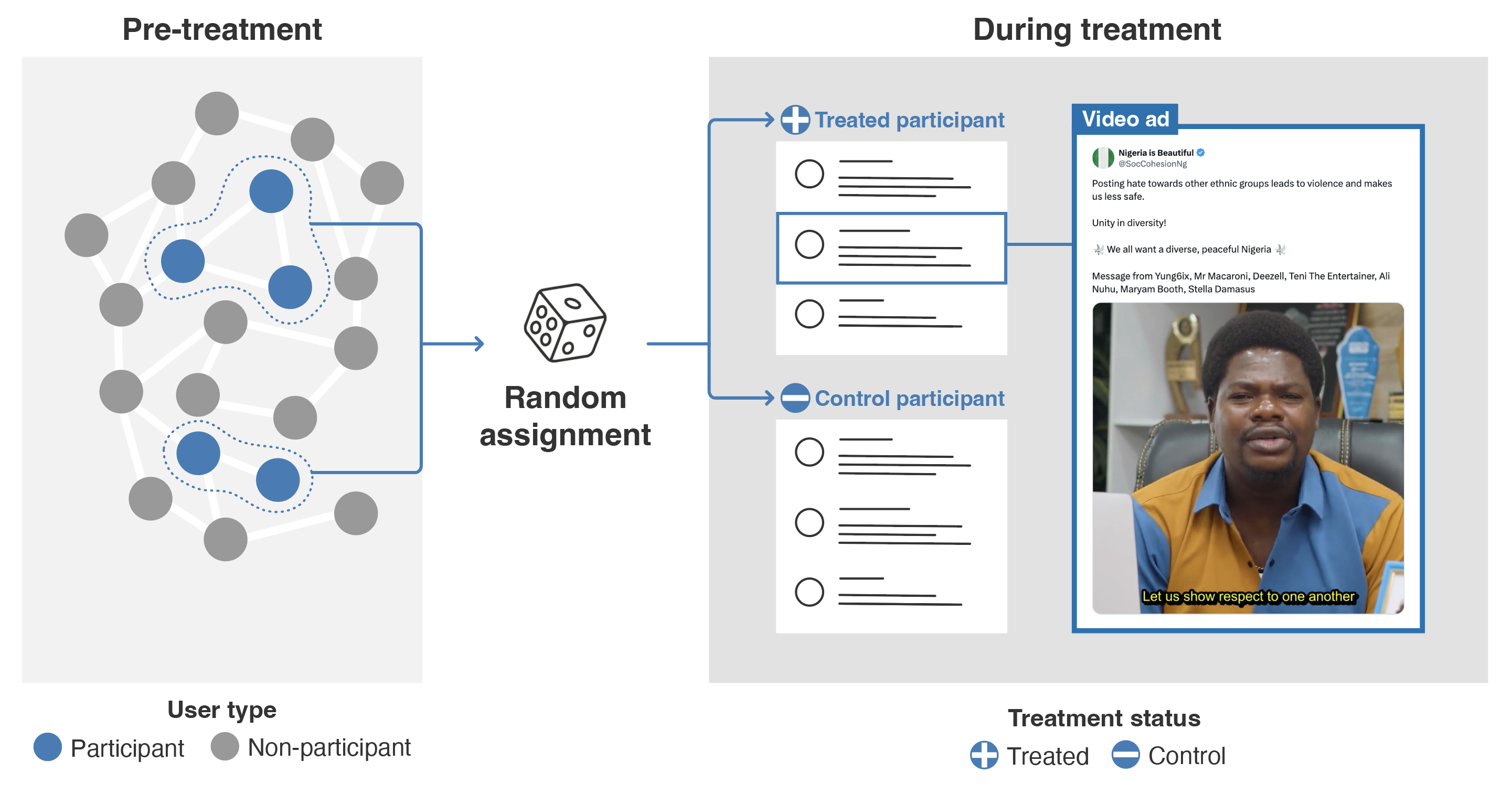}
    \caption{\textbf{Experimental sample and treatment delivery}. The left panel represents the \finalTotalUserN{} sampled X accounts that had engaged with hate speech in blue, with other accounts in grey. Users were grouped into clusters based on pre-treatment mention ties, and clusters were randomly assigned to treatment or control. The right panel shows that between November 29, 2023 and April 17, 2024, users assigned to treatment (plus sign) were eligible to receive prosocial video ads from Nigerian celebrities directly in their feed via the X Ads platform, whereas control users (minus sign) were not.}
    \label{fig:experiment_design}
\end{figure}

\begin{figure}[t]
  \centering
  \includegraphics[width=\linewidth]{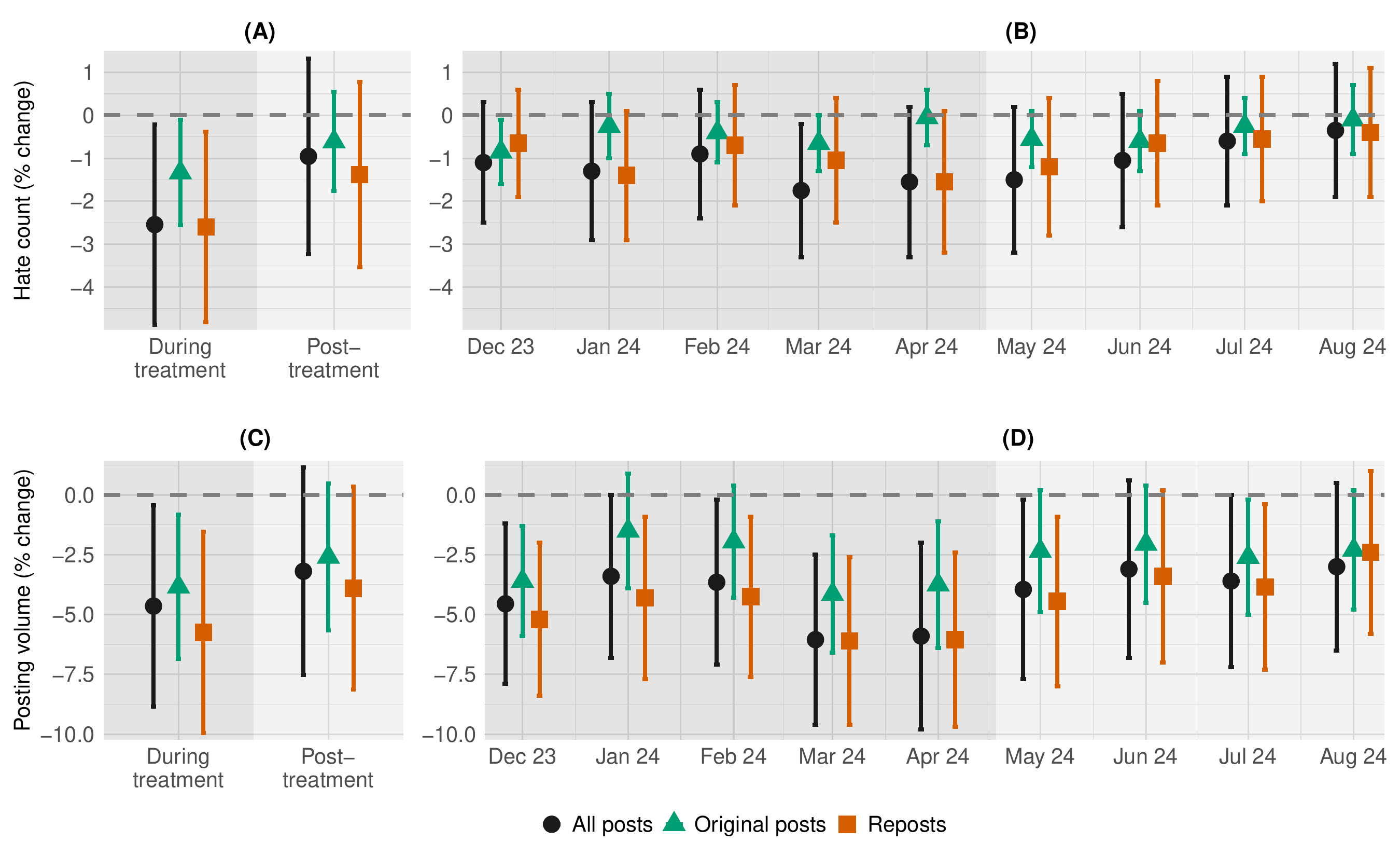}
  \caption{\textbf{Direct effects on hate posts and posting volume}. Panels A–B report intent-to-treat percentage changes in hate posts during the campaign treatment and post-treatment periods (A) and by month (B). Panels C–D show percentage changes in posting volume. Results are plotted separately for all posts (black), original posts (green), and reposts (red). Error bars show 95\% confidence intervals. Treated users reduced hate posting and overall activity during the campaign, with a large share of the reduction persisting post-treatment.}
  \label{fig:direct_effects}
\end{figure}

\begin{figure}[t]
  \centering
  \includegraphics[width=\linewidth]{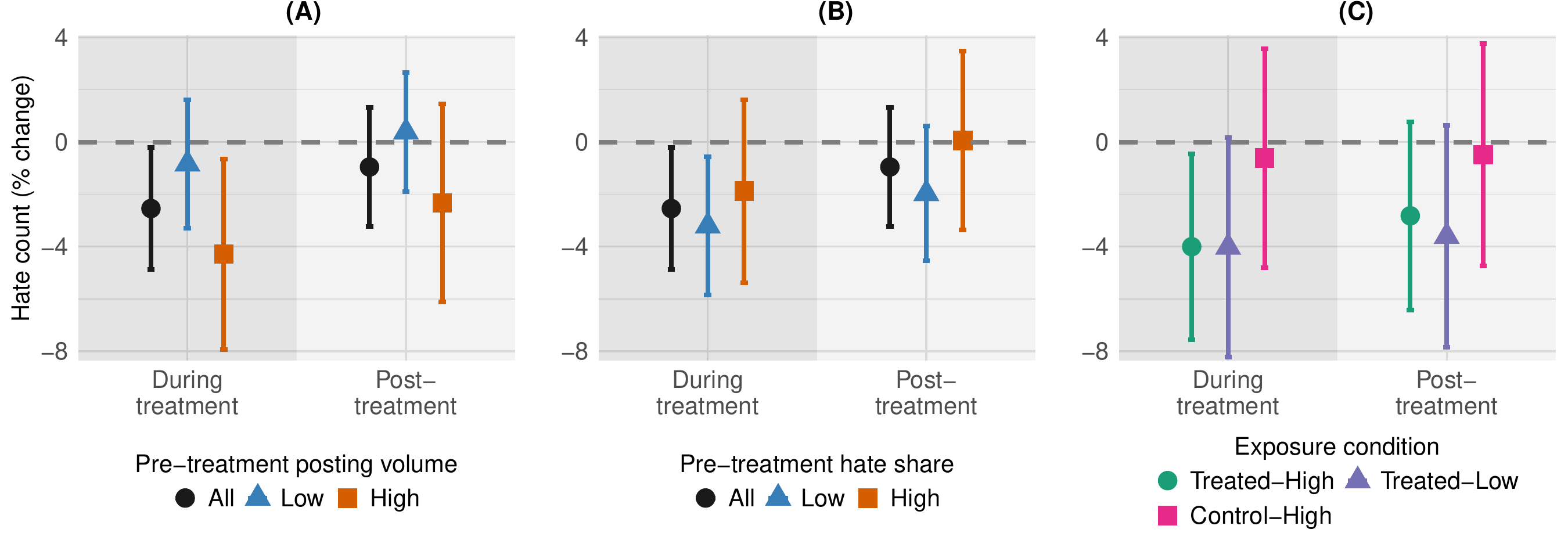}
  \caption{\textbf{Heterogeneous treatment effects on hate posts}. Treatment effects differed by pre-treatment posting volume (A), pre-treatment hate share (B), and exposure condition (C). Low and high groups for posting volume and hate share are based on median splits. Exposure combines own assignment with the share of peers assigned to treatment (see Methods). Error bars show 95\% confidence intervals. Effects were concentrated among more active and less hateful users, and there is no evidence of indirect effects among participants.}
  \label{fig:heterogeneity_direct}
\end{figure}

We analyzed over 1.7 billion public posts generated before July 31, 2023, using a language model fine-tuned on Nigerian text to detect hate content \citep{tonneau-etal-2024-naijahate}. High-score posts typically reflected explicit animosity (``My hate for northern people keeps growing'') or dehumanization (``Biafrans are animals''). We identified \finalTotalUserN{} users who had engaged with hate content targeting Nigerian groups at least twice between January 1, 2022 and July 31, 2023 (see Methods). These users were partitioned into clusters using a graph-based algorithm that grouped accounts frequently mentioning one another during this period \citep{ugander2013graph}. We then randomized clusters in equal proportions to treatment and control, with treated users eligible to receive preventive short videos through the X Ads platform (Fig. \ref{fig:experiment_design}; see Methods). Over the 140-day treatment period, the platform delivered ads to 17,650 users, representing 45.8\% of the treatment group. Ads were shown more than 900,000 times and generated substantial engagement, including likes and reposts (see Supplementary Information Section \ref{sec:si_experiment_design}).

We first assess whether the campaign reduced the production and sharing of hate content. Our preregistered primary outcomes are each user’s number of hate posts, computed separately for all posts, original posts, and reposts across three periods: pre-treatment (January 1 to July 31, 2023), treatment (November 29, 2023 to April 17, 2024), and post-treatment (May 1 to August 31, 2024). We estimate intent-to-treat (ITT) effects by comparing treated and control users, adjusting for pre-treatment levels and accounting for the cluster-randomized design (see Methods). Treated users reduced both the production and sharing of hate content during the campaign (Fig. \ref{fig:direct_effects}). Overall hate posts declined by 2.5\% relative to control ($p=0.032$), including a 1.3\% reduction in original hate posts ($p=0.033$) and 2.6\% reduction in hate reposts ($p=0.021$; Fig. \ref{fig:direct_effects}A-B). Because only 45.8\% of treated users received at least one ad, the ITT understates the effect of exposure. Using treatment assignment as an instrument for ad receipt, we estimate a treatment-on-the-treated effect of 5.5\%, consistent with prior evidence that digital messaging campaigns typically shift behavior by 1–5\% \citep{Athey2023digital, Gordon2022ads, lin2024reducing}. Importantly, about 75\% ($p<10^{-6}$) of the effect observed during treatment persisted into the post-treatment period (see Supplementary Information Section \ref{sec:si_persistence_direct}). Results are robust to alternative hate definitions and pre-treatment adjustments (see Supplementary Information Section \ref{sec:si_robustness_direct}).

To further characterize treatment effects, we examined non-preregistered outcomes beginning with posting volume. The campaign reduced total posting during treatment by 4.6\% ($p=0.03$), with $88$\% of the effect persisting post-treatment ($p<10^{-6}$), indicating a broader drop in engagement among treated users (Fig. \ref{fig:direct_effects}C-D). The decline in activity exceeded the reduction in hate content, although the difference was not significant ($p>0.32$). We also observe a shift in the accounts that treated users amplified. Despite making $5.7$\% fewer reposts overall ($p=0.007$), treated users redirected their attention toward more visible sources: the average account they reposted had 3.4\% more followers ($p=0.045$) and 2.8\% more posts ($p=0.051$) than accounts reposted by control users. This pattern suggests that the intervention shifted attention toward more prominent accounts even as total reposting declined (see Supplementary Information Section \ref{sec:si_composition_reposted}).

Treatment effects varied by pre-treatment behavior. Among users whose pre-treatment posting volume was above the median, hate posts declined by 4.3\% during treatment ($p=0.021$), whereas effects for those below the median were not significant ($p=0.50$; Fig. \ref{fig:heterogeneity_direct}A). Treatment effects also varied by pre-treatment hatefulness: among users whose pre-treatment share of hate content was below the median, hate posts declined by 3.2\% ($p=0.017$), while effects for those above the median were not significant ($p=0.29$; Fig. \ref{fig:heterogeneity_direct}B). Overall, the campaign was more effective among users who were more active and who posted less hate pre-treatment. Similar heterogeneity patterns emerged when total posting is used as the outcome (see Supplementary Information Section \ref{sec:si_heterogeneity_direct}).

We then test preregistered indirect effects within participants’ social networks. As part of the original randomized design, exogenous variation in local exposure to treated peers was generated by independently flipping a subset of individual assignments within each cluster, which shifted the share of treated neighbors (see Methods). This created four exposure conditions defined by a user’s own assignment and the share of their neighbors assigned to treatment: treated–high exposure, treated–low exposure, control–high exposure, and control–low exposure. Detecting indirect effects requires sustained interaction among participants, yet only 3.39\% of the accounts they reposted were in the experimental sample, and the set of reposted accounts changed by roughly 78\% month to month, leaving little stability through which indirect influence could occur (See Supplementary Information Section \ref{sec:si_turnover}). Consistent with this constraint, we detect no differences between control–high and control–low exposure ($p=0.77$) and no differences between treated–high and treated–low exposure relative to control–low exposure ($p=0.99$; see Fig. \ref{fig:heterogeneity_direct}C and Supplementary Information Section \ref{sec:si_indirect_participants}).

\begin{figure}[t!]
    \centering
    \includegraphics[width=0.8\linewidth]{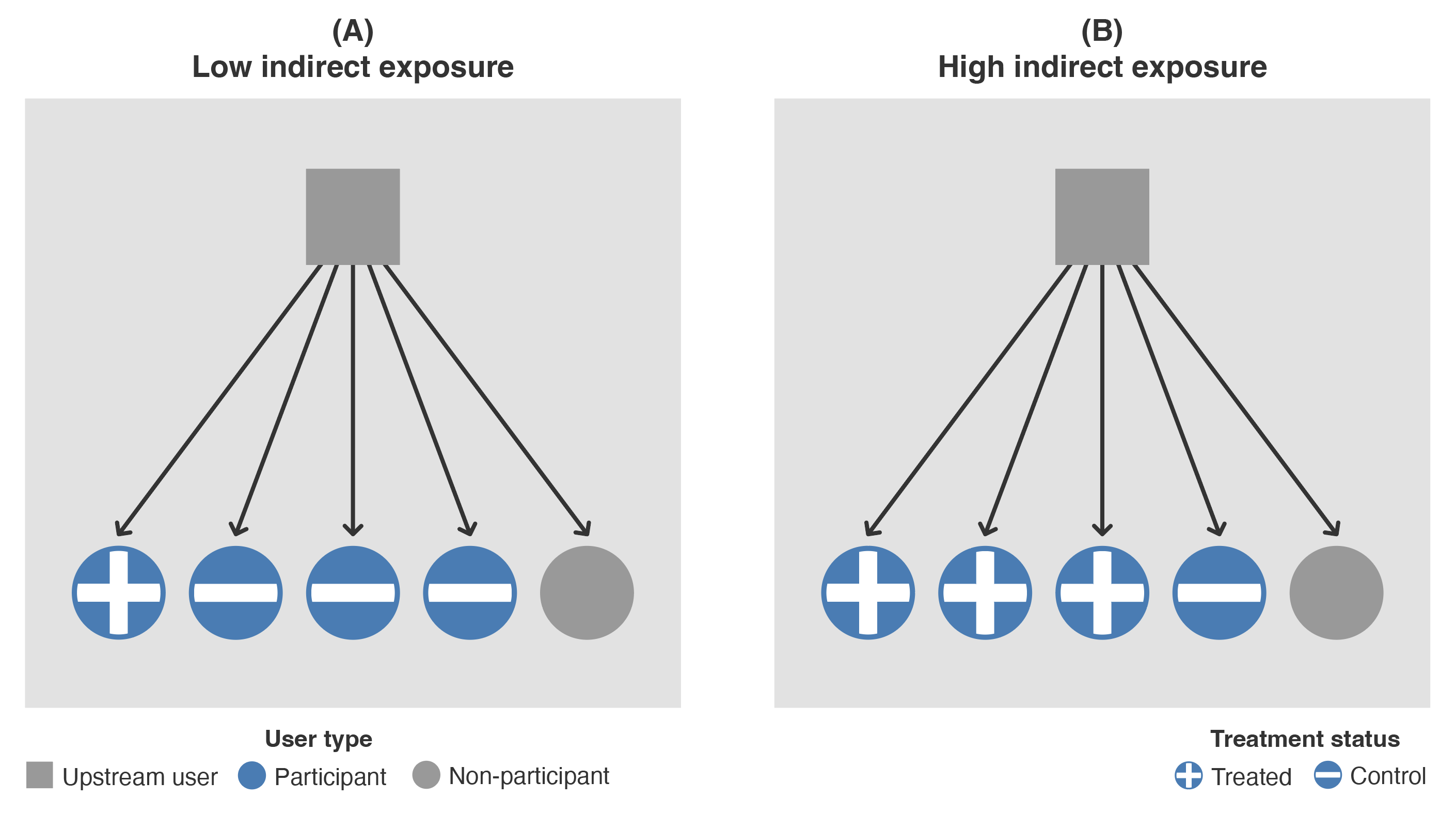}
    \caption{\textbf{Upstream users and indirect exposure}. Upstream users are accounts whose posts were frequently reposted by participants before treatment. (A) An upstream user has few treated participants among their reposters. (B) An upstream user with a larger share of treated reposters, implying greater indirect exposure to the intervention. Grey squares represent upstream users; blue circles represent participants; plus and minus symbols denote treated and control status; grey circles represent non-participants; arrows represent pre-treatment repost links.}
    \label{fig:upstream_design}
\end{figure}

\begin{figure}[ht!]
    \centering
        \includegraphics[width=0.8\linewidth]{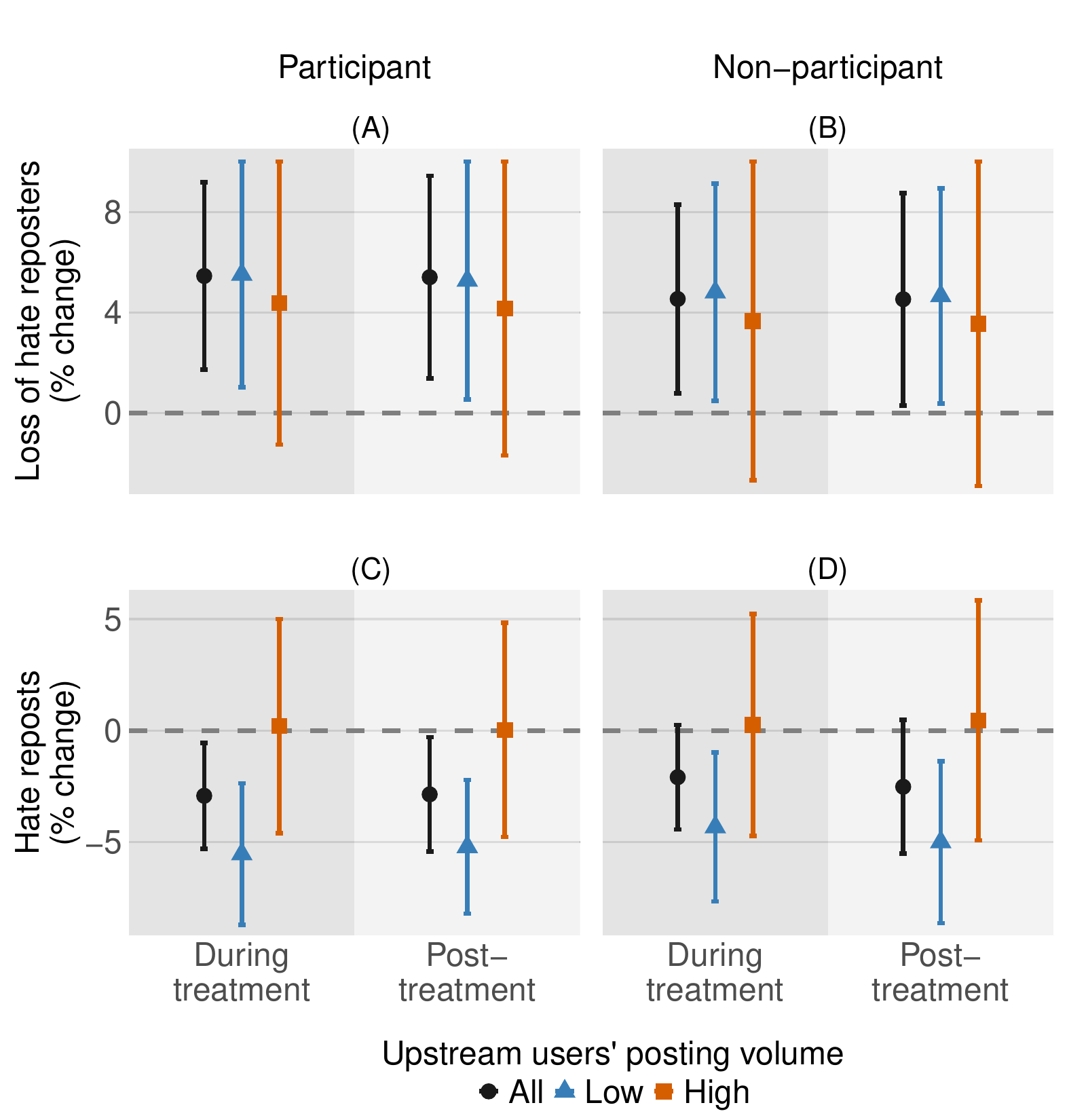}
    \caption{\textbf{Indirect effects on upstream users}. Upstream users are non-participants whose posts were frequently reposted by participants before treatment. Panels report percentage changes associated with a one-percentage-point increase in the share of an upstream user’s pre-treatment audience that was treated. Rows indicate outcome type (loss of pre-treatment hate reposters; hate reposts). Columns indicate audience type (participants; non-participants). Colors denote upstream users’ pre-treatment posting volume (all, low, high). Larger treated audience shares are associated with greater loss of hate reposters and fewer hate reposts, including among non-participants. Error bars show 95\% confidence intervals.}
    \label{fig:producers_heterogeneity}
\end{figure}

We then examined a setting with frequent user interaction, which increases the scope for indirect effects. Specifically, we tested for indirect effects on non-participants exposed through changes in what participants chose to repost (Fig. \ref{fig:upstream_design}). Following our preregistration, we focus on 400 upstream users, defined as non-participants whose content was frequently reposted by participants pre-treatment (see Methods). Since reposting is a primary amplification channel, reduced reposting by treated users can reduce an upstream user’s visibility and limit the downstream spread of hate content. We quantify indirect exposure by the share of each upstream user’s pre-treatment audience assigned to treatment, and estimate how a one-percentage-point increase in that share affects the loss of pre-treatment hate reposters and hate reposts (Fig. \ref{fig:producers_heterogeneity}).

Greater indirect exposure increased the loss of pre-treatment hate reposters by 5.4\% among participants ($p=0.013$) and 4.5\% among non-participants ($p=0.025$), with similar magnitudes persisting after the campaign ended (Fig. \ref{fig:producers_heterogeneity}A-B; Supplementary Information Section \ref{sec:si_persistence_indirect}). These increases were concentrated among upstream users with below-median posting volume, whereas effects for high-volume upstream users were not significant. We observe similar patterns for hate reposting (Fig. \ref{fig:producers_heterogeneity}C-D): a one-percentage-point increase in treated audience share reduced hate reposts by 2.9\% among participants ($p=0.019$) and by 2.1\% among non-participants ($p=0.075$). Applying this marginal effect to observed exposure levels implies an average reduction of roughly 53\% in upstream users' hate reposts. These patterns indicate that weakened amplification by treated users propagated to non-participants who never received the messages, limiting downstream spread beyond the treated population. Results are robust to alternative specifications (see Supplementary Information Section \ref{sec:si_robustness_indirect_upstream}). 

\section*{Discussion} \label{sec:discussion}

Here we show that a large-scale campaign of short prosocial videos from Nigerian celebrities reduced the production and sharing of hate content on X. Implemented in a context of persistent inter-ethnic tensions, the intervention produced sustained reductions in hateful posts and reposts that continued after the campaign ended. The effects also extended to users who never received messages, demonstrating the feasibility of proactive, non-censoring interventions.

Prior field experiments on direct counterspeech report reductions in hateful posts, but such approaches require real-time detection and are difficult to scale. By contrast, our preventive strategy based on targeted messaging reaches users before harmful content is posted. The magnitudes of our effects are comparable to those reported in recent digital messaging experiments in political, health, and misinformation contexts \citep{Athey2023digital, Gordon2022ads, lin2024reducing}. This suggests that proactive messaging can meaningfully curb hate while remaining compatible with large-scale deployment.

The limited and unstable overlap among the accounts that participants reposted over time left few opportunities for treatment to diffuse through local connections. This pattern likely reflects structural limits on local influence in social media platforms, which increasingly function as broadcast networks that elevate large accounts rather than facilitate repeated interaction among peers \citep{chen2023future_social_media}. Nevertheless, when a substantial share of a user’s audience was exposed to the intervention, amplification of that user’s hate content declined. This finding is especially relevant for users who produce higher levels of hate and are less responsive to direct treatment, because reducing engagement among their audiences can indirectly limit their reach \citep{gennaro2025hate}.

The indirect effects we observe among non-participants are unlikely to reflect simple interruption of repost chains. If treated users were disrupting repost chains, we would expect systematic changes in the relative rankings of non-participants within upstream users' repost audiences, but we observe no such pattern (see Supplementary Information Section \ref{sec:si_rank_reposte}).
Instead, the evidence is consistent with a ranking response inside the platform in which reduced engagement lowers the visibility of smaller upstream users.

These results highlight the promise of proactive messaging strategies as a complement to content moderation. Preventive campaigns can be delivered at low marginal cost through existing ad infrastructure, in contrast to manual review and enforcement systems that require substantial ongoing investment (see Supplementary Information Section \ref{sec:si_cost_effectiveness}). Future work should compare different messenger types and message framings \citep{munger2021don, hangartner2021empathy, siegel2020no2sectarianism}, and explore adaptive or AI-driven personalization \citep{bar2024generative, costello2024durably}. As platforms scale back moderation and algorithmic systems continue to amplify harmful content \citep{corsi2024evaluating, milli2025amplification}, preventive counterspeech offers a scalable and non-censoring approach that can reduce online harms and promote safer digital environments.

\section*{Materials and Methods} 
\label{sec:methods}

\paragraph{Preregistration.} The study was preregistered before data analysis. The experimental design, treatment assignment, and primary outcomes for participants were preregistered at \url{https://osf.io/q735x}. The analysis plan for indirect effects on upstream users was preregistered at \url{https://osf.io/rpygk}.

\paragraph{Sample selection.}
Between January 1, 2021 and December 31, 2022, we used the X API to identify users listing Nigeria as their profile location. We recursively added all Nigerian accounts mentioned in their posts, repeating this procedure until no additional users were identified. This resulted in more than 2.8 million users and over 1.7 billion posts, capturing a substantial share of the platform's national activity. Hate content was detected using NaijaXLM-T, a large language model fine-tuned on Nigerian social media text \citep{tonneau-etal-2024-naijahate}, which assigns each post a continuous hate score between 0 and 1 and classifies the targeted group, including ethnic, religious, and gender-based categories. Potential participants were X accounts among these users that had posted, reposted, or liked at least two posts classified as hate content targeting Nigerian ethnic groups between January 1, 2022 and July 31, 2023. We excluded inaccessible accounts and a small number of highly connected users to enable estimation of indirect effects, resulting in 80,154 experimental participants (see Supplementary Information Section \ref{sec:si_experiment_design} and \ref{sec:si_descriptive_stats}). \noActivityBadStatusUserN{} accounts became inactive before treatment or became inaccessible during or after the treatment period, yielding \finalTotalUserN{} users in the analysis sample. Attrition did not differ significantly between treatment and control groups (see Supplementary Information Section \ref{sec:si_attrition_participants}).

\paragraph{Treatment assignment.}
As described in our preregistration, we constructed an interaction network based on user mentions from January 1, 2022 to July 31, 2023, and partitioned participants into 4,253 clusters using a 3-net clustering algorithm \citep{ugander2013graph}. Treatment followed a two-stage graph-cluster randomization. First, clusters were independently assigned to treatment or control with equal probability. Second, each participant’s assignment was independently flipped with probability 0.18 to generate variation in within-cluster exposure \citep{eckles2017design}. Participants were stratified into ten equal-sized groups by pre-treatment activity level and uploaded to the X Ads platform, enabling us to examine ad-delivery patterns across activity levels. Prosocial video ads featuring Nigerian celebrities were delivered only to treated users. To prevent cross-arm contamination, ads were automatically deleted and reposted after any interaction, ensuring that engagements did not expose ads to control users (see Supplementary Information Section \ref{sec:si_intervention_delivery}).

\paragraph{Outcome measures.}
Following our preregistration, the primary outcomes are participants’ hate posts from January 1, 2023 to August 31, 2024. Hate content was measured using NaijaXLM-T, which assigns each post a continuous hate score between 0 and 1. For each user, we counted posts with a hate score above 0.5 and constructed separate measures for all posts, original posts, and reposts. Outcomes were log-transformed and averaged within each period for estimation. Robustness checks using alternative hate thresholds and specifications are reported in the Supplementary Information Section \ref{sec:si_robustness_direct}.

\paragraph{Estimation of direct effects on participants.}
We estimate treatment effects by comparing outcomes for treated and control participants, under the assumption of no indirect effects within the participant population. The average treatment effect is estimated using a difference estimator that adjusts each user’s outcome using a rescaled pre-treatment baseline to improve precision while preserving unbiasedness. Inference accounts for the graph-cluster randomization by analytically deriving the estimator’s variance under the joint treatment probabilities induced by the assignment procedure. Formal definitions and variance derivations are provided in Supplementary Information Section \ref{sec:si_direct_effect}.

\paragraph{Estimation of indirect effects on participants.}
To estimate indirect effects on participants, we assigned each user to one of four exposure conditions based on their own treatment assignment and the pre-treatment share of mention ties to treated participants. Exposure was defined using a $q$-fraction rule with $q = 0.7$, classifying users as having high exposure when at least 70\% of their pre-treatment mention ties to participants are with treated users, and low exposure when at least 70\% are with control users; users whose neighbor composition lies between these thresholds are not assigned to either exposure condition \citep{ugander2013graph}. This yields four conditions: treated–high exposure, treated–low exposure, control–high exposure, and control–low exposure. Exposure propensities under the graph-cluster randomization were obtained by Monte Carlo simulation of the assignment procedure and used to construct inverse-propensity weights. Exposure-specific means and their contrasts were then estimated using a weighted regression equivalent to the Hájek estimator, incorporating pre-treatment posting volume to improve precision. Variances were computed using the network-robust HAC estimator of \citet{gao2025causal}. Full methodological details are provided in Supplementary Information Section \ref{sec:si_indirect_participants}.

\paragraph{Upstream user selection.}
We identified accounts that participants frequently reposted during the four months preceding the campaign (August 1, 2023 to November 29, 2023) and ranked them by the share of their reposts coming from participants during this pre-treatment period. We excluded users with more than 15,000 posts prior to the campaign in 2023. The top 400 remaining accounts, each receiving at least 37\% of their reposts from participants, were selected as upstream users with substantial potential exposure. For these users, we collected all original posts between May 1, 2023 and August 31, 2024.

\paragraph{Estimation of indirect effects on upstream users.}
We estimate indirect effects on upstream users using randomization inference under the sharp null hypothesis that their outcomes are unaffected by variation in indirect exposure. For each upstream user $i$, exposure $T_i$ is defined as the share of their pre–treatment reposters who were assigned to treatment. Our test statistic is the coefficient on $T_i$ from a regression of each outcome on exposure, controlling for pre–treatment posting volume. We then simulate 10{,}000 reshuffled assignments that replicate the original experimental design, recompute exposures, and re–estimate the same regression to generate the distribution of the test statistic under the sharp null. Randomization inference $p$–values compare the observed statistic to this distribution, and confidence intervals for the marginal effect of a one–percentage–point increase in exposure are obtained by inverting the procedure. Further methodological details are provided in Supplementary Information Section \ref{sec:si_indirect_upstream}.

\singlespacing
\section*{Acknowledgments} This work was supported by funding from the United Kingdom’s Foreign, Commonwealth \& Development Office (FCDO) and the World Bank’s Research Support Budget. Additional support was provided through NYU IT High Performance Computing resources, services, and staff expertise. The findings, interpretations, and conclusions expressed in this article are those of the authors. They do not necessarily represent the views of the International Bank for Reconstruction and Development/World Bank and its affiliated organizations or those of the Executive Directors of the World Bank or the governments they represent. S.F. and V.O. are employees of the World Bank. E.J., B.K., M.T., H.L., and D.B. were consultants at the World Bank during the study. E.J. was supported by NSF grant \#1745640. H.L. is supported by Canadian Social Sciences \& Humanities Research Council Tri-Agency Funding (\#192324).
This study was reviewed and approved by the Health Media Lab Institutional Review Board (IRB \#2310).

\setlength\bibsep{0pt}
\bibliographystyle{abbrvnat}
\bibliography{references}
\clearpage

\renewcommand{\appendixname}{Supplementary Information}
\renewcommand{\appendixpagename}{Supplementary Information}
\renewcommand{\appendixtocname}{Supplementary Information}

% --- prevent the "Supplementary Information" block heading from appearing in the TOC ---
\addtocontents{toc}{\protect\setcounter{tocdepth}{-10}}
\begin{appendices}
    % re-enable TOC entries for SI sections/subsections
    \addtocontents{toc}
    {\protect\setcounter{tocdepth}{2}}
    %%%%%%%%%%%%%%%%%%%%%%%%%%
%% New numbering for appendix (using Prefix S for sections, figures, and tables)
\appendix
\renewcommand{\thesection}{S\arabic{section}}

\setcounter{figure}{0}
\renewcommand{\thefigure}{S\arabic{figure}}

\setcounter{table}{0}
\renewcommand{\thetable}{S\arabic{table}}
%%%%%%%%%%%%%%%%%%%%%%%%%%

\thispagestyle{empty}
\vspace{1em}

% add a bit more space between headings numbers and titles in TOC
\makeatletter
\@ifundefined{cftsecnumwidth}{%
  % Fallback (no tocloft): widen the number box used by \l@section in the TOC
  \renewcommand*\l@section{\@dottedtocline{1}{1.5em}{2em}}
    \renewcommand*\l@subsection{\@dottedtocline{2}{3.0em}{2.6em}}
}{%
  % If tocloft is loaded: widen the number width for sections/subsections
  \setlength{\cftsecnumwidth}{2em}
    \setlength{\cftsubsecnumwidth}{2.6em}
}
\makeatother
\tableofcontents
\clearpage

\section{Experiment design}
\label{sec:si_experiment_design}

\subsection{Hate speech detection}
\label{sec:si_hate_detection}
Hate content was identified using NaijaXLM-T, a large language model trained on Nigerian social-media text and described in \cite{tonneau-etal-2024-naijahate}. The model assigns each post a continuous hate score between 0 and 1 and is publicly available at \url{https://huggingface.co/worldbank/naija-xlm-twitter-base-hate}. Although the classifier detects multiple forms of hate speech, our intervention focused on ethnic hate. To identify targeted groups, we further fine-tuned the same model to predict whether a post targeted one of the eight Nigerian communities defined in \cite{tonneau-etal-2024-naijahate}. Users were eligible for inclusion in the experiment if, during the detection period from January 1, 2022 to July 31, 2023, they posted, reposted, or liked at least two posts with a hate score above 0.9 and an ethnic target score above 0.5. Manual review of a random sample of 50 eligible users indicated that approximately 80\% had clearly engaged in ethnic hate.

\subsection{Participant selection}
\label{sec:si_hateful_users}
Applying these eligibility criteria to the full dataset yielded 95{,}140 users who had engaged in ethnic hate between January 1, 2022 and July 31, 2023. We excluded 5{,}346 inaccessible accounts and removed the top 2.5\% of users by outgoing mentions and the top 2.5\% by incoming mentions during the same period to facilitate estimation of indirect effects, eliminating 3{,}956 highly active users. Among the remaining users, 80{,}154 had at least one repost or mention tie to another eligible user during the pre-treatment period and formed the experimental population; 5{,}684 isolated users were excluded. Of these participants, \noActivityUserN{} became inactive before treatment began between August 1 and November 29, 2023, and a further \badStatusUserN{} deactivated, were suspended, or became private during follow-up, yielding a final analysis sample of \finalTotalUserN{} participants.

\subsection{Interaction network among participants}
\label{si:network_construction}
We constructed a weighted directed network among participants using all interactions marked with the ``@'' symbol between January 1, 2022 and July 31, 2023. These interactions include mentions and reposts; for simplicity, we refer to all of them as mentions. A directed edge from user $a$ to user $b$ was added whenever $a$ mentioned $b$, with the edge weight equal to the total number of mentions. To reduce noise from infrequent ties, we retained only each user’s ten strongest outgoing and ten strongest incoming edges. The resulting pruned network, containing 521{,}961 edges, preserved the main channels of influence and served as the basis for clustering and treatment assignment.

\subsection{Network clustering}
Because users may be influenced by the treatment status of others, we account for interference by randomizing treatment at the level of clusters rather than individuals \citep{ugander2013graph}. Clusters were formed from the pre-treatment mention network, under the assumption that indirect effects operate primarily through direct ties. Following the 3-net clustering procedure proposed by \citet{ugander2013graph}, we selected cluster centroids iteratively, removed all users within two network hops of each centroid, and assigned remaining users to the nearest centroid. Our only deviation from the original procedure was in resolving ties: when users were equidistant from multiple centroids, we assigned them based on interaction intensity so that strongly connected users were grouped together. This approach reflects the heavy-tailed structure of online interactions and improves precision when estimating indirect effects.

\subsection{Cluster randomization}
\label{sec:si_treatment_randomization}
Treatment assignment followed a two-stage design based on graph-cluster randomization \citep{ugander2013graph, eckles2017design}. In the first stage, entire clusters were independently assigned to treatment or control with equal probability, so that all users within a cluster shared the same assignment. Cluster-level randomization alone would not allow separate identification of direct and indirect effects, since some users and all of their neighbors would receive the same assignment. To ensure that all users have a non-zero probability of being in any exposure condition, we independently flipped each user’s assignment with probability of $0.18$, a procedure often referred to as hole punching \citep{eckles2017design}. This second stage ensures that some users were surrounded by many treated neighbors and others by few, allowing estimation of both direct and indirect effects under the assumption of local interference. Cluster assignments were independent across clusters, and the resulting within-cluster treatment correlation was approximately 0.41, providing sufficient variation and power to estimate indirect effects.

\subsection{Prosocial messaging intervention}
The treatment consisted of a short video in which Nigerian celebrities conveyed a prosocial message. The celebrities represented diverse ethnic and gender backgrounds: Teni, a female Yoruba singer and songwriter; Maryam Booth, a female Hausa actress in Kannywood and Nollywood; Mr Macaroni, a male Yoruba comedian and content creator; Stella Damasus, a female Igbo actress and producer in Nollywood; Deezel, a Hausa male hip-hop artist and film producer; Ali Nuhu, a male Hausa actor in Kannywood and Nollywood, a producer and director in Kannywood and the managing director of the Nigerian Film Cooperation; Yung6ix, an Edo male rapper and hip-hop artist. The message they delivered was: \\
\emph{``Hello fellow Nigerians. Hi, this is Teni, Maryam Booth, Mr Macaroni, Stella Damasus, Deezel, Ali Nuhu, Yung6ix aka Swaggalomo on a spaceship. Please, let us stop posting hate to other ethnic groups.  Posting hate towards other ethnic groups is not right. Posting hate towards other ethnic groups contributes to violence and makes us all less safe. Let us show respect to one another. Be respectful of one another. We all want a peaceful Nigeria, for you, me and our children. Peace and love.''}

\begin{figure}[t!]
    \centering
    \includegraphics[width=0.5\linewidth]{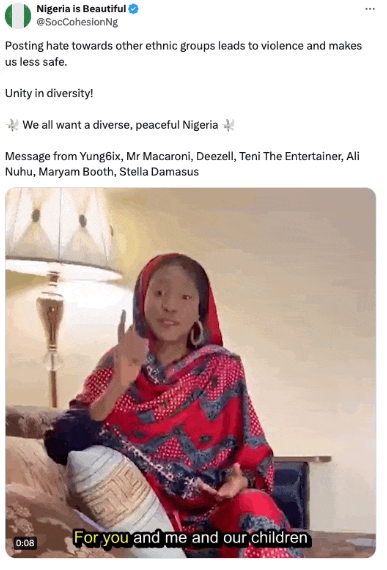}
    \caption{Screenshot of the ad appearing in treated user feeds}
    \label{fig:treatment_video}
\end{figure}

The video appeared in the feeds of treated participants, embedded in a promoted post as shown in Fig. \ref{fig:treatment_video}. There were seven versions of the video, all containing the same message but differing in the order in which celebrities were shown. A version of the video can be found under \url{https://drive.google.com/file/d/1AFNOHCL0A6Xfljs2hCL_00o3KWeesCLT/view?usp=sharing}. 

\subsection{Intervention delivery}
\label{sec:si_intervention_delivery}
All treated users were eligible to receive the video messages through the X Ads platform, but delivery depended on platform algorithms. In total, 17{,}650 users, representing 45.8\% of the treatment group, were served at least one ad. Across all ads served, the campaign generated 916{,}278 impressions and 44{,}872 engagements, including 17{,}795 link clicks, 5{,}679 likes, and 1{,}313 reposts. Treated users also mentioned the featured celebrities together with ad-related terms more often than controls. Mentions paired with keywords such as ``hate'' and ``violence'' increased by 0.08\% ($p=0.00047$), with a significant rise in their share of total posts ($p=0.03$), indicating that the ads attracted user attention. To prevent unintended exposure through social interactions, an ad was withdrawn from a user’s feed immediately after that user liked or reposted it and could later reappear only for that user. This protocol did not affect treatment assignment or eligibility for others. Because ads were withdrawn following each engagement, these figures likely understate engagement under standard delivery conditions.

\subsection{Descriptive statistics}
\label{sec:si_descriptive_stats}
Baseline activity, hate content, and network interactions during the pre-treatment period are reported in Table \ref{tab:si_summary_stats_pre}. Participants display wide variation in posting behavior, with most activity consisting of reposts rather than original posts. Hate posts (hate score $>$ 0.5) represent a small share of overall activity on average, though the distribution is highly skewed. Interactions with other participants account for only a small fraction of total interactions, indicating limited direct overlap in audiences. Covariate balance between treatment and control groups is reported in Table \ref{tab:si_covariate_balance_pre}. Mean differences are small for all variables, and none are statistically significant (all $p$-values $>0.2$), indicating that randomization produced comparable groups at baseline.

\begin{table}[t]
\centering
\caption{Descriptive statistics for the \finalTotalUserN{} participants during the pre-treatment period. Statistics are computed across users. Hate posts are defined as posts with a hate score greater than 0.5. Shares are expressed as percentages.}
\label{tab:si_summary_stats_pre}
\begin{tabular}{lcccc}
\toprule
Variable & Mean & SD & Median & IQR (25--75\%) \\
\midrule
\multicolumn{5}{l}{\textbf{Monthly posts}} \\
\quad Total posts & 417 & 910 & 133 & [36, 412] \\
\quad Original posts & 100 & 198 & 32 & [6.6, 108] \\
\quad Reposts & 318 & 853 & 52 & [7.6, 253] \\
\midrule
\multicolumn{5}{l}{\textbf{Monthly hate (score $>$ 0.5)}} \\
\quad Hate posts          & 3.9  & 11  & 0.86 & [0.14, 3.29] \\
\quad Original hate posts & 0.88 & 3.6 & 0.14 & [0.00, 0.57] \\
\quad Hate reposts        & 3.10 & 9.8 & 0.43 & [0.00, 2.14] \\
\quad Hate share (\%)     & 1.5  & 3.1 & 0.5  & [0.1, 1.4] \\
\midrule
\multicolumn{5}{l}{\textbf{Network interactions}} \\
\quad Followers & 873 & 2855 & 273 & [87, 826] \\
\quad Reposts of participants (\%) & 4.0 & 7.9 & 2.2 & [0.5, 4.7] \\
\bottomrule
\end{tabular}
\end{table}

\begin{table}[t]
\centering
\caption{Pre-treatment covariate balance between treatment and control groups.  All means and differences are in log-scale except hate share and the share of participant reposts, which are expressed in percentages. Two-sided $p$-values test equality of means and account for the randomization design. Hate posts are defined as posts with a hate score greater than 0.5. Shares are expressed as percentages.}
\label{tab:si_covariate_balance_pre}
\begin{tabular}{lcccc}
\toprule
Variable & Mean (Treatment) & Mean (Control) & Difference & $p$-value \\
\midrule
\multicolumn{5}{l}{\textbf{Monthly posts}} \\
\quad Total posts    & 4.73 & 4.76 & -0.00366 & 0.816 \\
\quad Original posts & 3.34 & 3.35 & -0.00111 & 0.916 \\
\quad Reposts        & 3.83 & 3.86 & -0.00289 & 0.828 \\
\midrule
\multicolumn{5}{l}{\textbf{Monthly hate (score $>$ 0.5)}} \\
\quad Hate posts          & 0.940  & 0.921  & 0.00190  & 0.613 \\
\quad Original hate posts & 0.348  & 0.341  & 0.000696 & 0.566 \\
\quad Hate reposts        & 0.728  & 0.710  & 0.00175  & 0.591 \\
\quad Hate share (\%)     & 1.76   & 1.69   & 0.0686   & 0.472 \\
\midrule
\multicolumn{5}{l}{\textbf{Network interactions}} \\
\quad Followers & 5.53 & 5.57 & -0.0391 & 0.823 \\
\quad Reposts of participants (\%) & 4.16 & 4.21 & -0.0446 & 0.757 \\
\bottomrule
\end{tabular}
\end{table}

\section{Direct effect estimation}
\label{sec:si_direct_effect}

\subsection{Conceptual approach}
This section describes the estimand, estimator, and inference procedures used to quantify the direct effect of the intervention on participants. We assume no interference, so that each user’s outcome depends only on their own treatment status. Under this assumption, standard individual-level estimators apply. Let $Y'_i(t)$ denote user $i$'s outcome under treatment status $t \in \{0,1\}$, aggregated by month or study period. Because most outcomes are counts, we analyze log-transformed outcomes:

\begin{align*}
Y_i(t) = \log\!\big(Y'_i(t) + 1\big).
\end{align*}

Our primary estimand is the average treatment effect (ATE):

\begin{align*}
\tau \;=\; \frac{1}{N}\sum_{i=1}^N \big(Y_i(1)-Y_i(0)\big).
\end{align*}

We report ATE estimates for multiple log-scale outcomes in Fig.~\ref{fig:direct_effects}. Because these estimates use a Horvitz--Thompson inverse-propensity weighted estimator, we first characterize the treatment propensities implied by the randomization design.

\subsection{Propensities}

Let $p_i(t)=\Pr(Z_i=t)$ denote the probability that unit $i$ receives treatment status $t\in\{0,1\}$, and let $p_{ij}(t_1,t_2)$ denote the joint probability that units $i$ and $j$ receive treatment statuses $t_1$ and $t_2$. Joint propensities are required for the variance expressions derived later. Under the graph-cluster randomization with hole punching described in Section~\ref{sec:si_treatment_randomization}, all units share identical marginal propensities, and assignments are independent across clusters.

\begin{proposition}
\label{prop:hp_propensities}
Under the design described in Section~\ref{sec:si_treatment_randomization}, the marginal propensity of each unit is

\begin{align*}
p_i(1) = p_t + p_{hp} - 2 p_t p_{hp}, 
\qquad
p_i(0) = 1 - p_t - p_{hp} + 2 p_t p_{hp},
\end{align*}

where $p_t$ is the probability that a cluster is assigned to treatment and $p_{hp}$ is the probability that an individual assignment is flipped. For distinct units $i \neq j$, let $c(i)$ denote the cluster of unit $i$. The joint propensities are:

\begin{align*}
p_{ij}(t_1,t_2)=
\begin{cases}
p_i(t_1)p_j(t_2), & c(i) \neq c(j), \\[0.75em]
p_t + p_{hp}^2 - 2p_t p_{hp}, & c(i)=c(j),~ t_1=t_2=1, \\[0.75em]
1 + 2p_t p_{hp} + p_{hp}^2 - 2p_{hp} - p_t, & c(i)=c(j),~ t_1=t_2=0, \\[0.75em]
p_{hp}(1-p_{hp}), & c(i)=c(j),~ t_1 \neq t_2.
\end{cases}
\end{align*}
For $i=j$, we recover the marginal propensity: $p_{ii}(t,t)=p_i(t)$ and $p_{ii}(t_1,t_2)=0$ if $t_1\neq t_2$.
\end{proposition}

In our experiment, the cluster-level treatment probability is $p_t = 0.5$ and the hole punching probability is $p_{hp} = 0.18$. Substituting these values yields $p_i(1)=p_i(0)=0.5$, and for distinct units in the same cluster:

\begin{align*}
p_{ij}(1,1)=p_{ij}(0,0)=0.352,
\qquad
p_{ij}(1,0)=p_{ij}(0,1)=0.148.
\end{align*}

These propensities are used in the inverse-probability weighting and in the variance calculations that follow.

\subsection{Point estimation}

Under the no-interference assumption and the randomization described in Section~\ref{sec:si_treatment_randomization}, the Horvitz--Thompson (HT) estimator for the mean potential outcome under treatment status $t$ is
\begin{align*}
\widehat{\mu}(t)
= \frac{1}{N} \sum_{i: Z_i=t} \frac{Y_i^{\text{during}}}{p_i(t)}.
\end{align*}

Because user activity is highly heterogeneous, we improve precision using a difference estimator \citep{aronow2013class}. Let $Y_i^{\text{pre}}$ denote the pre-treatment outcome, rescaled by a constant $\alpha$ so that the mean of $\alpha Y^{\text{pre}}$ matches the mean of $Y^{\text{during}}$. Define the adjusted outcome:

\begin{align*}
\Delta_i^{\text{during}} = Y_i^{\text{during}}-\alpha Y_i^{\text{pre}}.
\end{align*}

The HT difference estimator for the ATE is then
\begin{align}
\widehat{\tau}_{\text{diff}} = \widehat{\mu}_{\text{diff}}(1) - \widehat{\mu}_{\text{diff}}(0),
\label{eq:si_direct_effect_estimator}
\end{align}

where

\begin{align*}
\widehat{\mu}_{\text{diff}}(t) = \frac{1}{N} \sum_{i: Z_i=t} \frac{\Delta_i^{\text{during}}}{p_i(t)}.
\end{align*}

Because the pre-period outcome is independent of treatment, this estimator remains unbiased while achieving substantially lower variance \citep{aronow2013class}. All direct effect estimates reported in the main text and SI are based on $\widehat{\tau}_{\text{diff}}$.

\subsection{Inference under graph-cluster randomization}
Valid inference requires accounting for correlation in assignment within clusters. The variance of $\widehat{\tau}_{\text{diff}}$ is obtained from the variances and covariance of the HT mean estimators under graph-cluster randomization, using the marginal and joint propensities $p(t)$ and $p(t_1,t_2)$ derived above. For compactness, all quantities refer to the differenced outcomes. We have the following theorem, which characterizes the variance and covariance of $\widehat{\mu}(t)$.

\begin{theorem}[Variance and covariance under graph-cluster randomization]
The variance and covariance of the Horvitz--Thompson estimators $\widehat{\mu}(t)$ under graph-cluster randomization with hole punching are given by
\begin{align}
\mathrm{Var}\big[\widehat{\mu}(t)\big]
= \frac{1}{N^2} \left[\left(\frac{p(t,t)}{p(t)^2}-1\right)\sum_{c \in \mathcal{C}} S_c(t,t)
+ \left(\frac{p(t)-p(t,t)}{p(t)^2}\right) Q(t,t)\right]
\label{eq:var_diff_potential_outcome}
\end{align}
for $t \in \{0,1\}$, and for $t_1 \neq t_2$
\begin{align}
\mathrm{Cov}\big[\widehat{\mu}(t_1),\widehat{\mu}(t_2)\big]
= \frac{1}{N^2} \left[\left(\frac{p(t_1,t_2)}{p(t_1)p(t_2)}-1\right)\sum_{c \in \mathcal{C}} S_c(t_1,t_2)
- \frac{p(t_1,t_2)}{p(t_1)p(t_2)}\, Q(t_1,t_2)\right]
\label{eq:cov_diff_potential_outcome}
\end{align}
\end{theorem}

where $\mathcal{C}$ denotes the set of clusters, and 
\begin{align*}
S_c(t_1,t_2) &= \sum_{i \in c} \sum_{j \in c} Y_i(t_1)\,Y_j(t_2) \\
Q(t_1,t_2) &= \sum_{i=1}^N Y_i(t_1)\,Y_i(t_2)
\end{align*}

\begin{proof}
We rely on the general variance and covariance formulas for Horvitz--Thompson estimators in \citet{aronow2017estimating}. 
The variance of $\widehat{\mu}(t)$ after simplifying the notation in \citep{aronow2017estimating} can be written as 
\begin{align}
\mathrm{Var}\big[\widehat{\mu}(t)\big]
= \frac{1}{N^2} \sum_{i=1}^N \sum_{j=1}^N \left(\frac{p_{ij}(t,t)}{p_i(t)p_j(t)} - 1\right) Y_i(t)\,Y_j(t)
\label{eq:ht_variance_general}
\end{align}
where $p_{ij}(t,t) = \Pr(Z_i=t, Z_j=t)$ is the joint propensity for units $i$ and $j$. Under the hole punching design, all units share the same marginal propensity $p_i(t) = p(t)$, all pairs $i \neq j$ within the same (different) cluster(s) share the same joint propensity $p_{ij}(t,t)$. We use this to split the double sum in \eqref{eq:ht_variance_general} into within-cluster pairs and diagonal components. First, note that for any pair $(i,j)$ belonging to \emph{different} clusters, $p_{ij}(t,t) = p(t)^2$ so the factor in parentheses is
\begin{align*}
\frac{p_{ij}(t,t)}{p_i(t)p_j(t)} - 1 = \frac{p(t)^2}{p(t)^2} - 1 = 0,
\end{align*}
and these terms drop out. Thus, only pairs within the same cluster contribute. For $i \neq j$ in the same cluster $c$, we have $p_{ij}(t,t) = p(t,t)$, so
\begin{align*}
\frac{p_{ij}(t,t)}{p_i(t)p_j(t)} - 1 = \frac{p(t,t)}{p(t)^2} - 1
\end{align*}
which is constant over such pairs. For $i=j$, we have $p_{ii}(t,t) = p(t)$ and hence
\begin{align*}
\frac{p_{ii}(t,t)}{p_i(t)p_i(t)} - 1 = \frac{p(t)}{p(t)^2} - 1 = \frac{1}{p(t)} - 1
\end{align*}

Using these observations, we rewrite \eqref{eq:ht_variance_general} as
\begin{align*}
\mathrm{Var}\big[\widehat{\mu}(t)\big]
&= \frac{1}{N^2} \left[
\left(\frac{p(t,t)}{p(t)^2}-1\right)
\sum_{c \in \mathcal{C}} \sum_{\substack{i \in c \\ j \in c \\ i \neq j}} Y_i(t)\,Y_j(t)
+
\left(\frac{1}{p(t)}-1\right)
\sum_{i=1}^N \big(Y_i(t)\big)^2
\right]
\end{align*}

By definition, the within-cluster off-diagonal sum is
\begin{align}
\sum_{\substack{c \in \mathcal{C},\, i \in c,\, j \in c \\ i \neq j}} Y_i(t_1)\,Y_j(t_2)
= \sum_{c \in \mathcal{C}} S_c(t_1,t_2) - Q(t_1,t_2) 
\label{eq:si_y_ij_decomposition}
\end{align}

Substituting back gives
\begin{align*}
\mathrm{Var}\big[\widehat{\mu}(t)\big]
&= \frac{1}{N^2} \left[
\left(\frac{p(t,t)}{p(t)^2}-1\right) \big(\sum_{c \in \mathcal{C}} S_c(t,t) - Q(t,t)\big)
+
\left(\frac{1}{p(t)}-1\right) Q(t,t)
\right] \\
&= \frac{1}{N^2} \left[\left(\frac{p(t,t)}{p(t)^2}-1\right)\sum_{c \in \mathcal{C}} S_c(t,t)
+ \left(\frac{p(t)-p(t,t)}{p(t)^2}\right) Q(t,t)\right]
\end{align*}
which establishes \eqref{eq:var_diff_potential_outcome}. For the covariance between $\widehat{\mu}(t_1)$ and $\widehat{\mu}(t_2)$ with $t_1 \neq t_2$, \citet{aronow2017estimating} give
\begin{align}
\mathrm{Cov}\big[\widehat{\mu}(t_1),\widehat{\mu}(t_2)\big]
= \frac{1}{N^2} \sum_{i=1}^N \sum_{j=1}^N \left(\frac{p_{ij}(t_1,t_2)}{p_i(t_1)p_j(t_2)} - 1\right) Y_i(t_1)\,Y_j(t_2)
\label{eq:ht_cov_general}
\end{align}

Under our design, $p_{ii}(t_1,t_2) = 0$ for $t_1 \neq t_2$, so the diagonal terms become
\begin{align*}
\frac{p_{ii}(t_1,t_2)}{p_i(t_1)p_i(t_2)} - 1 = -1
\end{align*}
and for $i \neq j$ in the same cluster we have a common value $p_{ij}(t_1,t_2) = p(t_1,t_2)$. For $i \neq j$ in different clusters, $p_{ij}(t_1,t_2) = p(t_1)p(t_2)$, so those terms drop out as before. Splitting the sum into within-cluster and diagonal components yields
\begin{align*}
\mathrm{Cov}\big[\widehat{\mu}(t_1),\widehat{\mu}(t_2)\big]
&= \frac{1}{N^2} \left[
\left(\frac{p(t_1,t_2)}{p(t_1)p(t_2)}-1\right)
\sum_{\substack{c \in \mathcal{C},\, i \in c,\, j \in c \\ i \neq j}} Y_i(t_1)\,Y_j(t_2)
- \sum_{i=1}^N Y_i(t_1)\,Y_i(t_2)
\right]
\end{align*}

Using the decomposition \ref{eq:si_y_ij_decomposition} above, we can rewrite the covariance as
\begin{align*}
\mathrm{Cov}\big[\widehat{\mu}(t_1),\widehat{\mu}(t_2)\big]
&= \frac{1}{N^2} \left[
\left(\frac{p(t_1,t_2)}{p(t_1)p(t_2)}-1\right)
\big(\sum_{c \in \mathcal{C}} S_c(t_1,t_2) - Q(t_1,t_2)\big)
- Q(t_1,t_2)
\right] \\
&= \frac{1}{N^2} \left[
\left(\frac{p(t_1,t_2)}{p(t_1)p(t_2)}-1\right)\sum_{c \in \mathcal{C}} S_c(t_1,t_2)
- \frac{p(t_1,t_2)}{p(t_1)p(t_2)} Q(t_1,t_2)
\right]
\end{align*}
\end{proof}

Given the variance and covariance of estimated average potential outcomes \ref{eq:var_diff_potential_outcome} and \ref{eq:cov_diff_potential_outcome}, we can easily compute the variance of the estimated average treatment effect ($\widehat{\tau}$):
\begin{align}
\mathrm{Var}\!\big[\widehat{\tau}\big]
= \mathrm{Var}\!\big[\widehat{\mu}(1)\big]
+ \mathrm{Var}\!\big[\widehat{\mu}(0)\big]
- 2\,\mathrm{Cov}\!\big[\widehat{\mu}(1),\widehat{\mu}(0)\big]
\label{eq:var_diff}
\end{align}

\subsection{Conservative variance estimation}

We next construct variance and covariance estimators for the Horvitz--Thompson estimators $\widehat{\mu}(t)$. We follow \citet{aronow2017estimating}, specializing their results to our design where all units share the same marginal propensities $p_i(t)=p(t)$ and, for units in the same cluster, the same joint propensities $p_{ij}(t_1,t_2)=p(t_1,t_2)$.

\subsubsection{Variance estimator}

\citet{aronow2017estimating} propose the following design-based estimator for the variance of $\widehat{\mu}(t)$:
\begin{align*}
\widehat{\mathrm{Var}}\big[\widehat{\mu}(t)\big]
= \frac{1}{N^2} \sum_{i: Z_i=t} \sum_{j: Z_j=t}
\left( \frac{1}{p_i(t) p_j(t)} - \frac{1}{p_{ij}(t,t)} \right) Y_i Y_j
\end{align*}

Under our design, $p_i(t)=p(t)$ for all $i$ and $p_{ij}(t,t)=p(t,t)$ for all $i \neq j$ in the same cluster, while $p_{ij}(t,t) = p(t)^2$ for units in different clusters. Using the same cluster-level notation as in Theorem~\ref{eq:var_diff_potential_outcome}, this estimator simplifies to
\begin{align}
\widehat{\mathrm{Var}}\big[\widehat{\mu}(t)\big]
= \frac{1}{N^2} \left[
\left( \frac{1}{p(t)^2} - \frac{1}{p(t,t)} \right) \sum_{c \in \mathcal{C}} \widehat{S}_c(t,t)
+ \left( \frac{1}{p(t,t)} - \frac{1}{p(t)} \right) \widehat{Q}(t,t)
\right]
\label{eq:var_ht_estimator}
\end{align}
where
\begin{align*}
    \widehat{S}_c(t_1, t_2) &= \sum_{\substack{i: i \in c \\Z_i=t_1}} \sum_{\substack{j: j \in c\\Z_j=t_2}} Y_i Y_j \\
    \widehat{Q}(t) &= \sum_{i: Z_i=t} Y_i^2
\end{align*}

\citet{aronow2017estimating} show that, when all joint propensities $p_{ij}(t,t)$ are strictly positive for $i \neq j$, this variance estimator is unbiased:
\begin{align*}
\mathbb{E}\big[ \widehat{\mathrm{Var}}\big[\widehat{\mu}(t)\big] \big]
= \mathrm{Var}\big[\widehat{\mu}(t)\big]
\end{align*}
This condition is satisfied under the hole punching design, since every pair of units has positive probability of jointly receiving treatment $t$.

\subsubsection{Covariance estimator}

For the covariance between $\widehat{\mu}(t_1)$ and $\widehat{\mu}(t_2)$ with $t_1 \neq t_2$, \citet{aronow2017estimating} propose the following estimator:
\begin{align}
\begin{aligned}
\widehat{\mathrm{Cov}}\big[\widehat{\mu}(t_1),\widehat{\mu}(t_2)\big]
= \frac{1}{N^2} \Big[&
\sum_{i: Z_i=t_1} \sum_{\substack{j: Z_j=t_2 \\ j \neq i}}
\left( \frac{1}{p_i(t_1)p_j(t_2)} - \frac{1}{p_{ij}(t_1,t_2)} \right) Y_i Y_j \\
&\quad - \sum_{i: Z_i=t_1} \frac{Y_i^2}{2 p_i(t_1)}
- \sum_{i: Z_i=t_2} \frac{Y_i^2}{2 p_i(t_2)} \Big]
\end{aligned}
\label{eq:ht_cov_est_general}
\end{align}

Under our design, $p_i(t) = p(t)$ and $p_{ij}(t_1,t_2) = p(t_1,t_2)$ for $i \neq j$ in the same cluster, and $p_{ij}(t_1,t_2) = p(t_1)p(t_2)$ for units in different clusters. As before, only within-cluster pairs contribute to the first term. Using the cluster-level notation, the estimator simplifies to
\begin{align}
\begin{aligned}
\widehat{\mathrm{Cov}}\big[\widehat{\mu}(t_1),\widehat{\mu}(t_2)\big]
= \frac{1}{N^2} \Big[&
\left( \frac{1}{p(t_1)p(t_2)} - \frac{1}{p(t_1,t_2)} \right) \sum_{c \in \mathcal{C}} \widehat{S}_c(t_1,t_2) \\
&\quad - \frac{1}{2 p(t_1)} \widehat{Q}(t_1,t_1)
- \frac{1}{2 p(t_2)} \widehat{Q}(t_2,t_2) \Big]
\end{aligned}
\label{eq:ht_cov_est_simplified}
\end{align}

As shown in \citet{aronow2017estimating}, this covariance estimator is \emph{downward biased}, as long as $p_{ij}(t_1, t_2) > 0$ for $i \neq j$ and the outcomes $Y_i$ are non-negative. This condition is satisfied in our randomization design, thus
\begin{align*}
\mathbb{E}\big[ \widehat{\mathrm{Cov}}\big[\widehat{\mu}(t_1),\widehat{\mu}(t_2)\big] \big]
\leq \mathrm{Cov}\big[\widehat{\mu}(t_1),\widehat{\mu}(t_2)\big]
\end{align*}

\subsubsection{Conservative variance of the ATE}

Combining the variance and covariance estimators above yields a plug-in estimator for the variance of the ATE
\begin{align}
\widehat{\mathrm{Var}}\big[\widehat{\tau}_{\text{diff}}\big]
= \widehat{\mathrm{Var}}\big[\widehat{\mu}(1)\big]
+ \widehat{\mathrm{Var}}\big[\widehat{\mu}(0)\big]
- 2\,\widehat{\mathrm{Cov}}\big[\widehat{\mu}(1),\widehat{\mu}(0)\big]
\label{eq:var_tau_hat_conservative}
\end{align}

Because $\widehat{\mathrm{Var}}\big[\widehat{\mu}(t)\big]$ is unbiased for each $t$ and $\widehat{\mathrm{Cov}}\big[\widehat{\mu}(1),\widehat{\mu}(0)\big]$ underestimates the true covariance, the resulting estimator in \eqref{eq:var_tau_hat_conservative} is conservative:
\begin{align*}
\mathbb{E}\big[\widehat{\mathrm{Var}}\big[\widehat{\tau}_{\text{diff}}\big]\big]
\geq \mathrm{Var}\big[\widehat{\tau}_{\text{diff}}\big]
\end{align*}

All confidence intervals for direct effects reported in the supplementary information and the main text are based on this conservative variance estimator.

\subsection{Randomization inference}
As a complementary robustness check, we also report 95\% confidence intervals for monthly outcomes obtained by inverting randomization inference tests under the exact graph-cluster design. For each month, we compute the treatment coefficient from a regression of $Y_i^m$ on treatment with binned $Y_i^{\text{pre}}$ as a covariate, simulate 10{,}000 assignments from the hole punching design, and construct confidence intervals as the set of effects not rejected at the 5\% level.

\section{Persistence of direct effects}
\label{sec:si_persistence_direct}
A key question for our study is the extent to which the treatment effects observed during the intervention period persist after ads are no longer delivered. Rather than measuring persistence by comparing treatment and control groups directly in the post-treatment period, we follow the methodology introduced in \citet{lin2025persuading} to estimate the \emph{fraction of the treatment effect during the treatment period that remains in the post-treatment period}. This approach models persistence as a proportional relationship between immediate and delayed changes and has been shown to provide substantially greater statistical power and interpretability than comparing treatment groups at follow-up \citep{lin2025persuading}. In the next section, we briefly explain this methodology and present the results of the persistence parameter estimation.

\begin{table}[t]
\centering
\caption{Estimation and inference on the fraction of the during-treatment effect persisting into the post-treatment period for each outcome variable.}
\label{tab:si_persistence}
\begin{tabular}{lccc}
\toprule
Outcome & $\widehat{\beta}$ & SE (cluster-robust) & $p$-value \\
\midrule
Log hate posts            & 0.751 & 0.00376 & $< 10^{-6}$ \\
Log original hate posts   & 0.642 & 0.00430 & $< 10^{-6}$ \\
Log hate reposts          & 0.752 & 0.00422 & $< 10^{-6}$ \\
Log total posts           & 0.876 & 0.00273 & $< 10^{-6}$ \\
\bottomrule
\end{tabular}
\end{table}

\subsection{Conceptual approach}
\label{sec:si_persistence_method_direct_effect}
To formalize persistence, we begin by specifying outcome models for the treatment and post-treatment periods. Let $Y_i^{pre}, Y_i^{during}, Y_i^{post}$ denote user $i$'s pre-treatment, treatment and post-treatment outcomes, and let $Z_i \in \{0,1\}$ indicate treatment assignment. During the treatment period, we model outcomes as

\begin{align}
Y_i^{\text{during}} &= \alpha^{\text{during}} + \gamma^{\text{during}} \,Y_i^{\text{pre}} + \tau Z_i + \epsilon_i^{\text{during}},
\end{align}
where $\tau$ is the effect during the treatment period. The inclusion of $Y_i^{pre}$ adjusts for any linear trends in subsequent outcomes. Similarly, we model outcomes in the post-treatment period as
\begin{align*}
Y_i^{\text{post}} &= \alpha^{\text{post}} + \gamma^{\text{post}} \,Y_i^{\text{pre}} + \beta \tau Z_i + \epsilon_i^{\text{post}}.
\end{align*}
The coefficient $\beta$ is the fraction of the effect during treatment $\tau$ that persists in the post-treatment period, and it is our estimand of interest. $\beta = 1$ implies full persistence while $\beta = 0$ implies complete decay.
As in the during-treatment model, the term $\gamma^{\text{post}}\,Y_i^{\text{pre}}$ captures linear trends unrelated to persistence. 

While these structural models clarify the interpretation of $\beta$, estimating persistence directly from treatment indicators can be statistically inefficient, particularly when post-treatment effects are smaller. Instead, we can recover $\beta$ from a single regression model that relates the post-treatment outcome to the during-treatment outcome:
\begin{align}
Y_i^{\text{post}} = \alpha' + \gamma' Y_i^{\text{pre}} + \beta\,Y_i^{\text{during}} + \varepsilon_i'.
\label{eq:si_persistence_model_raw}
\end{align}

However, this specification may lack power because outcomes exhibit substantial variance. Following the same strategy used in our analysis and in \citet{lin2025persuading}, we instead work with changes relative to the pre-treatment period:
\begin{alignat*}{3}
& \Delta_i^{\text{during}} & \;=\; & Y_i^{\text{during}} - Y_i^{\text{pre}} \\
& \Delta_i^{\text{post}}   & \;=\; & Y_i^{\text{post}}   - Y_i^{\text{pre}}.
\end{alignat*}
We then estimate the persistence parameter using
\begin{align}
\Delta_i^{\text{post}} = \alpha' + \gamma' Y_i^{\text{pre}} + \beta\,\Delta_i^{\text{during}} + \varepsilon_i'.
\label{eq:si_persistence_model}
\end{align}

Simulations in \citet{lin2025persuading} demonstrated that this estimator consistently recovers the true persistence parameter and provides substantially greater statistical power than estimating post-treatment effects directly from treatment assignment.

\subsection{Results}
We estimated Equation~\eqref{eq:si_persistence_model} separately for the outcomes reported in the main text. Instead of using a single linear term for pre-treatment outcome, we divided its range into 40 equal-sized brackets and included them as dummy variables in the model. This better accounts for potential non-linearity in the trend. Because treatment in our experiment was assigned under a clustered randomization design, we used cluster-robust standard errors, clustering on the randomization clusters constructed by the 3-net algorithm.
Table~\ref{tab:si_persistence} reports the estimated persistence coefficients $\widehat{\beta}$, along with cluster-robust standard errors and $p$-values.

\begin{figure}[t]
    \centering
    \includegraphics[width=0.8\linewidth]{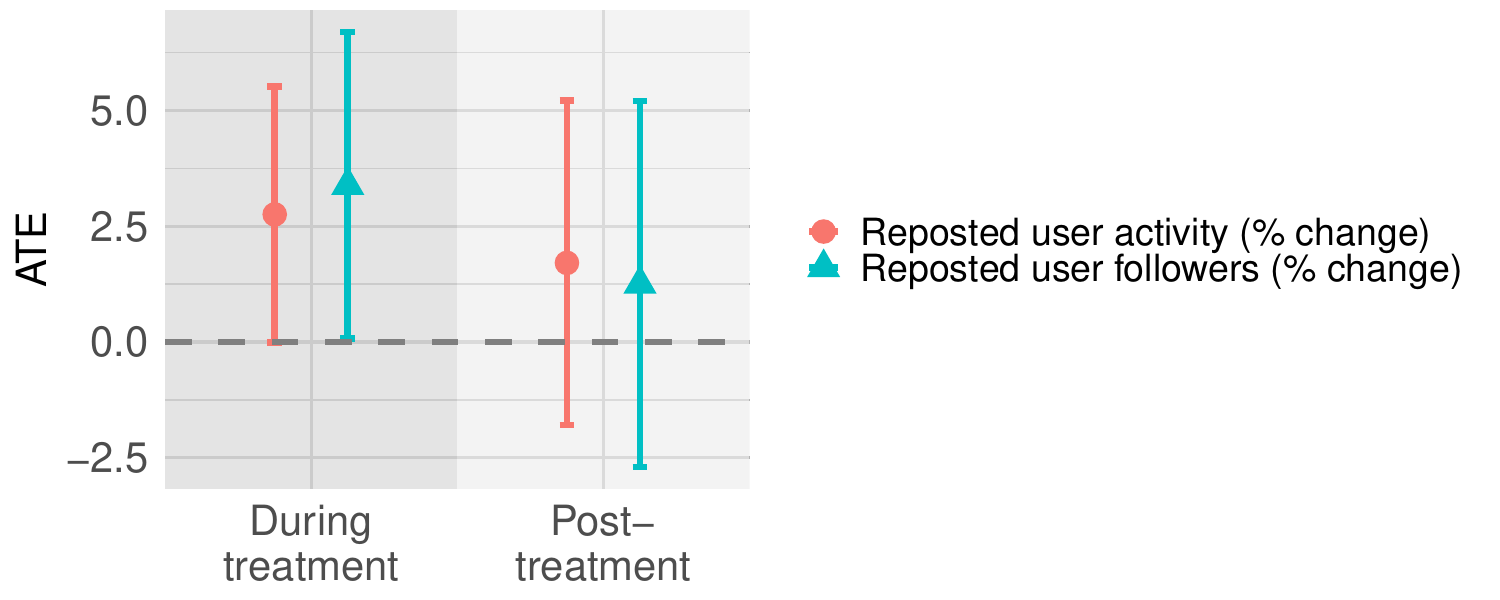}
    \caption{Composition of reposted accounts.
Estimated ATE on the average log follower count and average log status count of reposted accounts. Error bars correspond to 95\% confidence intervals.}
    \label{fig:original_user_follower_status}
\end{figure}

\section{Robustness of direct effects}
\label{sec:si_robustness_direct}

To assess the robustness of our main findings (Fig. \ref{fig:direct_effects}), we conducted several robustness checks. These checks validate that the estimated treatment effects are not sensitive to (i) the choice of pre-treatment adjustment period, or (ii) using alternative measures of hate. In addition to validating the study design, we conducted a negative control test over a period during which we do not expect any effect. Each robustness check replicates the estimation procedure described in the main text, using identical model specifications and inference under the experimental design.

\subsection{Alternative pre-treatment period}

\begin{figure}[t]
    \centering
    \begin{subfigure}[t]{\linewidth}
        \centering
        \includegraphics[width=\linewidth]{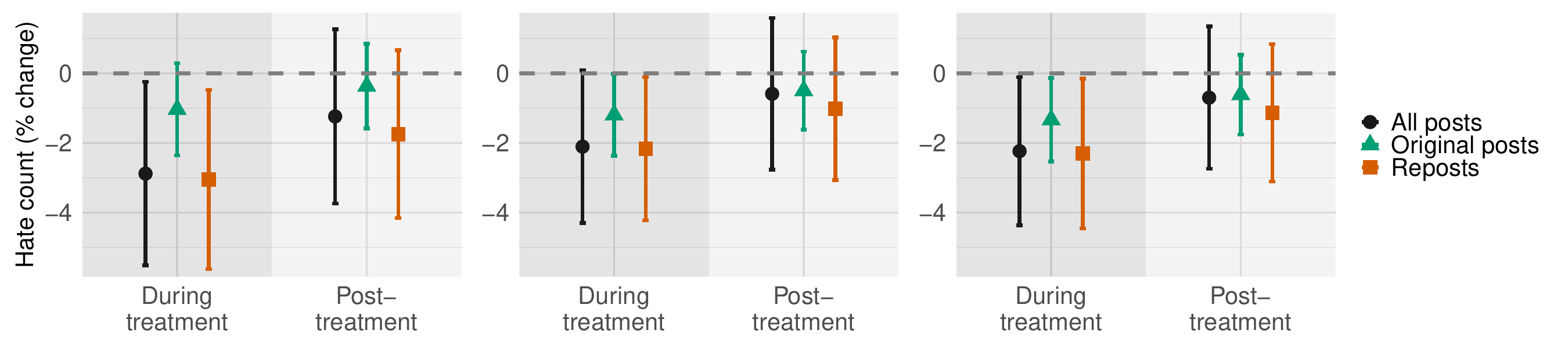}
        \caption{Hate outcomes \vspace{1.5em}}
    \end{subfigure}
    \begin{subfigure}[t]{\linewidth}
        \centering
        \includegraphics[width=\linewidth]{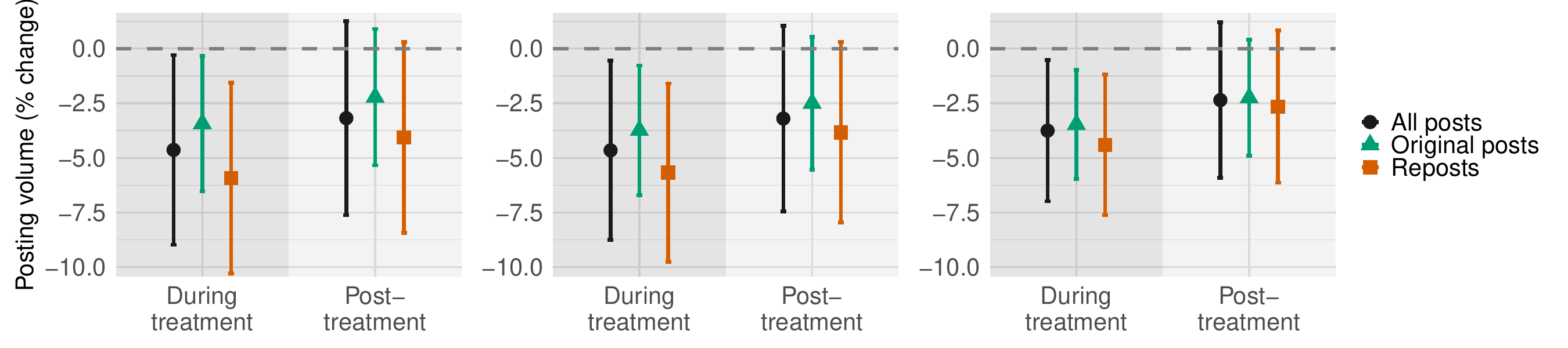}
        \caption{Posting volume outcomes}
    \end{subfigure}

    \caption{Robustness of treatment effects across alternative pre-treatment periods:
Jan 1–May 31, 2023 (left); Jan 1–Sept 30, 2023 (middle); May 1–Aug 31, 2023 (right). Error bars show 95\% confidence intervals.}
    \label{fig:robust_pretreat_all}
\end{figure}

To ensure that our results are not driven by the choice of pre-treatment period (January 1, 2023 to July 31, 2023), we re-estimated all the effects using three alternative baseline periods: (i) January 1, 2023 to September 30, 2023, (ii) January 1, 2023 to May 31, 2023 and (iii) May 1, 2023 to August 31, 2023. As detailed in Section~\ref{si_negative_control_test}, treated users deleted posts from the two months prior to treatment at a higher rate than control users after exposure. This suggests that October–November 2023, and to a lesser extent September 2023, may contain partial treatment leakage and are therefore excluded from the pre-treatment windows.

\begin{table}[t]
  \centering
  \caption{ATEs on different outcome variables by pre-treatment window.}
  \label{tab:robust_pretreat}
  \begin{tabular}{lcccc}
    \toprule
    Outcome &
    Jan--Sep~2023 &
    Jan--May~2023 &
    May--Aug~2023 &
    \makecell{Jan--Jul~2023 \\ (main text)} \\
    \midrule
    \midrule

    Total hate &
    \makecell{$-2.1\%$ \\ ($p=0.06$)} &
    \makecell{$-2.9\%$ \\ ($p=0.032$)} &
    \makecell{$-2.2\%$ \\ ($p=0.039$)} &
    \makecell{$-2.5\%$ \\ ($p=0.032$)} \\[6pt]

    Original hate &
    \makecell{$-1.2\%$ \\ ($p=0.046$)} &
    \makecell{$-1.0\%$ \\ ($p=0.126$)} &
    \makecell{$-1.3\%$ \\ ($p=0.029$)} &
    \makecell{$-1.3\%$ \\ ($p=0.033$)} \\[6pt]

    Reposted hate &
    \makecell{$-2.2\%$ \\ ($p=0.039$)} &
    \makecell{$-3.1\%$ \\ ($p=0.020$)} &
    \makecell{$-2.3\%$ \\ ($p=0.036$)} &
    \makecell{$-2.6\%$ \\ ($p=0.021$)} \\[6pt]

    \midrule
    \midrule
    
    Total posts &
    \makecell{$-4.7\%$ \\ ($p=0.026$)} &
    \makecell{$-4.6\%$ \\ ($p=0.036$)} &
    \makecell{$-3.8\%$ \\ ($p=0.022$)} &
    \makecell{$-4.6\%$ \\ ($p=0.030$)} \\
    
    Original posts &
    \makecell{$-3.7\%$ \\ ($p=0.013$)} &
    \makecell{$-3.4\%$ \\ ($p=0.030$)} &
    \makecell{$-3.5\%$ \\ ($p=0.006$)} &
    \makecell{$-3.8\%$ \\ ($p=0.013$)} \\
    
    Reposts &
    \makecell{$-5.7\%$ \\ ($p=0.006$)} &
    \makecell{$-5.9\%$ \\ ($p=0.008$)} &
    \makecell{$-4.4\%$ \\ ($p=0.007$)} &
    \makecell{$-5.7\%$ \\ ($p=0.007$)} \\

    \bottomrule
  \end{tabular}
\end{table}

For each specification, we repeated the difference estimator used in the main text to estimate the average treatment effect for all six log-transformed outcomes: total hate posts, original hate posts, reposted hate posts, total posts, original posts, and reposts. Fig.~\ref{fig:robust_pretreat_all} displays the estimated coefficients for these main outcomes across these pre-treatment definitions. Table~\ref{tab:robust_pretreat} reports the ATEs and their p-values for each outcome and pre-treatment window, including the one used in the main text.
The point estimates are consistent across all pre-treatment specifications and remain statistically significant at $p < 0.05$, with the exception of one.
These results indicate that the observed reductions in hate content are not artifacts of specific pre-treatment adjustment choices.

\subsection{Alternative measures of hate}

The main analysis defines hate as a binary indicator, summed across all posts, that equals one when the hate score assigned by NaijaXLM-T exceeds a threshold of 0.5. To assess sensitivity to this definition, we repeat the estimation using alternative binary thresholds, the continuous hate score, and a keyword-based measure.

\begin{figure}[t]
    \includegraphics[width=\linewidth]{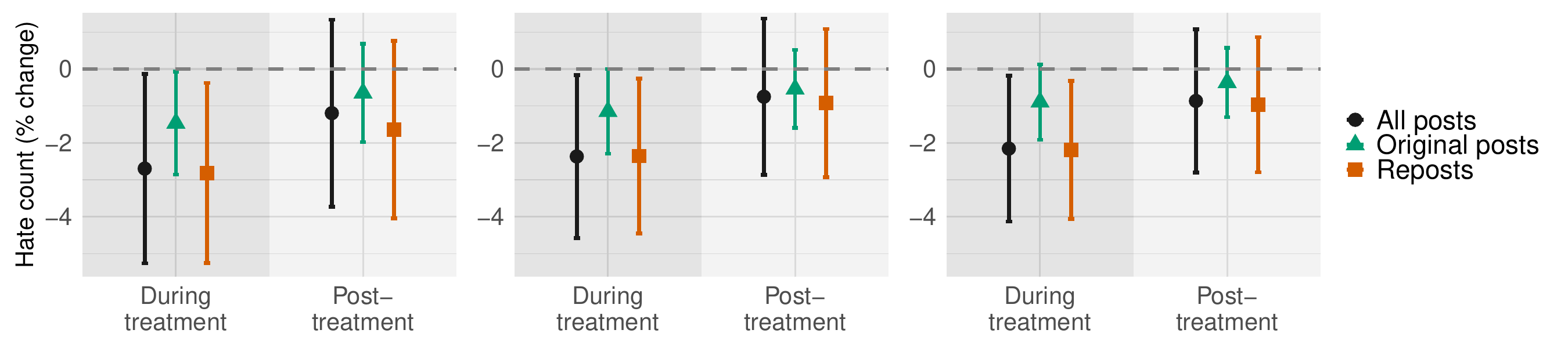}
    \caption{Robustness of treatment effects using an alternative outcome defined by the count of hate posts based on binary thresholds of the hate score: 0.1 (left), 0.75 (middle), and 0.9 (right).}
    \label{fig:si_robust_binary}
\end{figure}

\paragraph{Binary thresholds.}  
We modified the analysis of the main text by using different binary thresholds of 0.1, 0.75, or 0.9 on the model-assigned hate score. Fig.~\ref{fig:si_robust_binary} reports the resulting ATEs for each threshold. Across all thresholds, treatment effects remain negative and comparable in magnitude, demonstrating that our findings are not driven by the choice of cutoff.

\begin{figure}[t]
    \includegraphics[width=\linewidth]{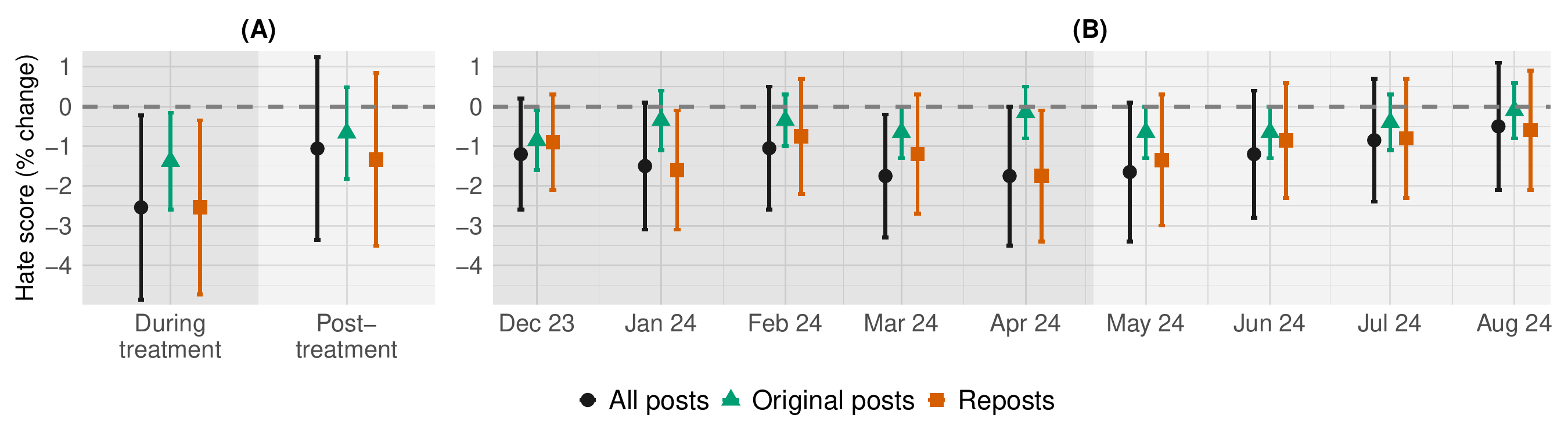}
    \caption{Robustness of treatment effects using hate score as the outcome, during- and post-treatment (A) and by month (B).}
    \label{fig:si_robust_score}
\end{figure}

\paragraph{Continuous score.}  
As an alternative to binarizing the hate score, we construct outcomes by directly summing the model-assigned hate scores across posts and applying the same log transformation used in the main analysis. This measure places greater weight on posts with higher predicted hate intensity while retaining information below any fixed threshold. Figure~\ref{fig:si_robust_score} reports the resulting period-level ATEs, along with monthly RI-based estimates, analogous to Fig.~\ref{fig:direct_effects}A and B in the main text. The estimated effects are consistent in sign and magnitude with those obtained using the binary hate measures.

\begin{figure}[t!]
    \includegraphics[width=\linewidth]{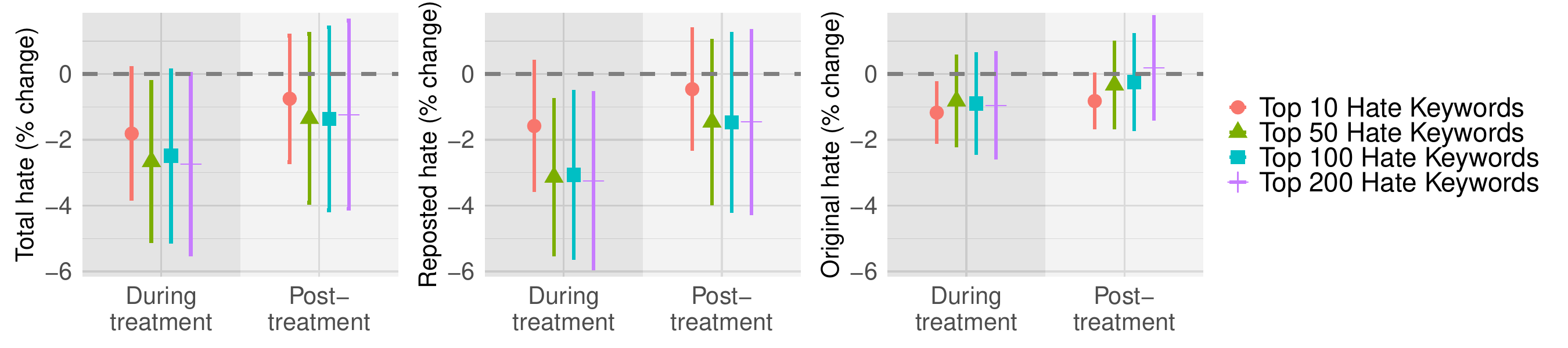}
    \caption{Robustness of treatment effects using an alternative outcome defined by the count of posts containing top keywords associated with hate. Panels show results for total hate posts (left), reposted hate posts (middle), and original hate posts (right).}
    \label{fig:si_robust_kw}
\end{figure}

\paragraph{Keyword-based counts.}  
As an additional robustness check, we construct an alternative measure of hate based on the presence of hate-related keywords. To select these keywords, we leveraged the annotated dataset in \cite{tonneau-etal-2024-naijahate} and computed the frequency of each word in both the hateful posts and the non-hateful posts. We then computed the ratio of these frequencies, measuring how much more likely a word is to appear in hateful posts than in non-hateful posts. Using this ranking, we identify the top $K$ hate-related keywords, with $K \in \{10, 50, 100, 200\}$, and measure hate as the count of posts containing at least one of these keywords. Fig.~\ref{fig:si_robust_kw} shows that the treatment effect remains negative, significant, and consistent across these definitions. The point estimates are larger than threshold-based binary measures.

Overall, robustness analyses using binary, score-based, and keyword-based measures confirm the stability of the treatment effects across alternative hate measures. Table~\ref{tab:si_robust_altdefs} reports the ATEs along with their p-values for all alternative definitions.

\begin{table}[t]
  \centering
  \caption{Estimated ATEs of log outcomes, expressed in terms of percentage change, and their p-values during treatment under alternative hate definitions.}
  \label{tab:si_robust_altdefs}
  \begin{tabular}{lccc}
    \toprule
    Definition & Total hate & Original hate & Reposted hate \\
    \midrule
    Score $\ge 0.1$  &    $-2.70, p=0.039$   &   $-1.46, p=0.040$ &   $-2.82, p=0.024$ \\
    Score $\ge 0.75$ &    $-2.37, p=0.035$   &   $-1.15, p=0.049$ &   $-2.35, p=0.028$ \\
    Score $\ge 0.9$  &    $-2.15, p=0.033$   &   $-0.89, p=0.086$ &   $-2.19, p=0.022$ \\
    Raw score        &    $-2.54, p=0.032$   &   $-1.38, p=0.027$ &   $-2.54, p=0.023$ \\
    Top-10 keywords  &    $-1.81, p=0.074$   &   $-1.17, p=0.010$ &   $-1.58, p=0.112$ \\
    Top-50 keywords  &    $-2.66, p=0.031$   &   $-0.82, p=0.236$ &   $-3.13, p=0.009$ \\
    Top-100 keywords &    $-2.49, p=0.061$   &   $-0.90, p=0.238$ &   $-3.06, p=0.018$ \\
    Top-200 keywords &    $-2.74, p=0.050$   &   $-0.96, p=0.239$ &   $-3.25, p=0.017$ \\
    \bottomrule
  \end{tabular}
\end{table}

\subsection{Negative control test}\label{si_negative_control_test}

\begin{figure}[t!]
    \includegraphics[width=\linewidth]{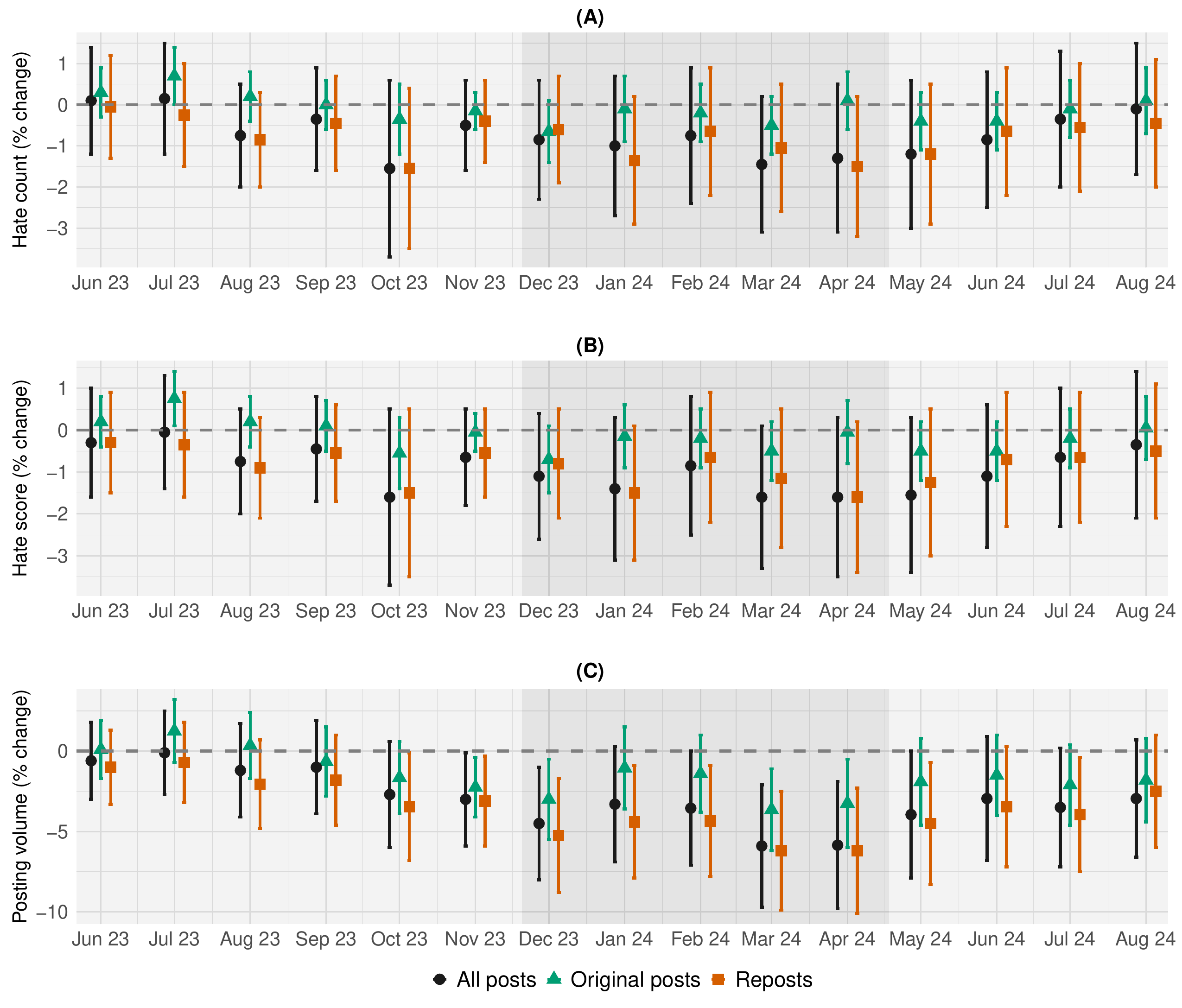}
    \caption{Negative control test based on monthly treatment effects estimated relative to a pre-treatment baseline. Treatment effects are estimated for the same outcomes as in the main text (binary hate counts and posting volume), as well as the continuous hate score. The first grey region indicates the placebo period, while the second and third grey regions represent the during- and post-treatment months, respectively. Error bars show 95\% randomization inference confidence intervals.}
    \label{fig:si_robust_negative_control}
\end{figure}

As an additional validation of our results, we conducted a negative-control analysis to assess differences between the treatment and control groups before the campaign. Detecting no systematic pre-treatment differences would strengthen confidence that the estimated effects during the intervention reflect the true causal impact of the ads rather than pre-existing trends.

We refer to the months immediately preceding the campaign as the \emph{placebo period}. The estimation procedure mirrors that used for the main analysis: outcomes measured during the placebo period are adjusted by a pre-treatment baseline using the same difference estimator described in the main text. For this test, the baseline adjustment period spans January 1 to June 1, 2023, leaving the period from June 1 to November 28, 2023 as the placebo period.

If the experimental design is valid, we would expect null or negligible effects during the placebo period, followed by stronger treatment effects during the intervention, and smaller effects in the post-treatment months. Fig.~\ref{fig:si_robust_negative_control}, which shows monthly treatment effects for binary hate count, hate score, and posting volume, confirms this pattern. The estimated monthly treatment effects during the placebo period are statistically indistinguishable from zero for at least the first four months, indicating that the observed treatment effects are not attributable to pre-existing behavioral differences between groups. However, some of the effects become slightly negative in the final two months before treatment. We interpret the late decline as selective content deletion among treated users who, after receiving the treatment, retrospectively removed some of their earlier posts from the recent past. Similar deletion behavior has been documented in prior social media experiments on direct counterspeech \cite{hangartner2021empathy}.

\subsection{Attrition among participants}
\label{sec:si_attrition_participants}

In this section, we assess whether the intervention induced differential attrition. For each outcome, we estimate the average treatment effect using the same Horvitz–Thompson difference estimator and conservative variance estimator defined in Equation~\ref{eq:var_tau_hat_conservative}. Across all measures, attrition rates do not differ significantly between treatment and control groups.

\subsubsection{Pre-treatment inactivity}
We first examine the users who became inactive before treatment rollout and remained inactive throughout the experiment.
For each user, we define an indicator of inactivity over the six months preceding treatment. Using the HT difference estimator and the conservative variance estimator in equation~\ref{eq:var_tau_hat_conservative}, we find that the estimated difference of inactivity rates between treatment and control is $-0.00197$ ($p=0.346$), indicating a slightly higher inactivity rate in the control group. This difference is not statistically significant, indicating no detectable imbalance in pre-treatment inactivity across treatment groups.

\subsubsection{Post-treatment inaccessibility}
Next, we assess whether users become inaccessible, either due to deletion, privatization, or suspension, at different rates across treatment groups. Applying the same estimation framework to the post-treatment inaccessibility indicator, we estimate an ATE of $-0.000724$ ($p=0.769$). This difference is not statistically significant, indicating that the intervention did not increase the likelihood of users becoming inaccessible after treatment.

\subsubsection{Combined attrition}
Finally, we consider a combined measure capturing whether a user either (i) became inactive before treatment or (ii) became inaccessible after treatment. Using the same estimator, the ATE for this combined attrition measure is $-0.00269$ ($p=0.458$). Again, the control group had a slightly higher attrition rate, but the difference between the treatment and control groups was not statistically significant.

Overall, the intervention did not induce differential attrition, suggesting that selective dropout is unlikely to bias the estimated treatment effects.

\section{Composition of reposted users}
\label{sec:si_composition_reposted}

As shown in Figure~\ref{fig:direct_effects}, treated users exhibit lower overall posting and reposting activity relative to control users. In this section, we examine whether the \emph{composition} of accounts that treated users repost differs from that of control users, conditional on a repost occurring.

To characterize the composition of reposted accounts, we record characteristics of the reposted account for each repost, focusing on (i) follower count and (ii) status count, defined as the total number of posts made by the account since inception. Because both variables are highly skewed, we analyze their logarithms.
For each user, we compute the average log follower count and the average log status count of the accounts they reposted pre-, during, and post-treatment, averaging across all reposts made by that user. These user-level averages serve as outcome variables. We then estimate the average treatment effect (ATE) of the intervention on each outcome, using the same estimation and inference procedures as in the direct effects analysis.

Figure~\ref{fig:original_user_follower_status} reports the estimated treatment effects. Despite a reduction in reposting volume, treated users tend to repost content from more popular and more active accounts when they do repost. Specifically, the accounts reposted by treated users have, on average, approximately 3.4\% more followers and 2.8\% higher lifetime posting activity than those reposted by control users. These results suggest that treatment shifts reposting behavior toward higher-profile accounts rather than uniformly reducing engagement.

\begin{figure}[t]
    \centering
    \includegraphics[width=\textwidth]{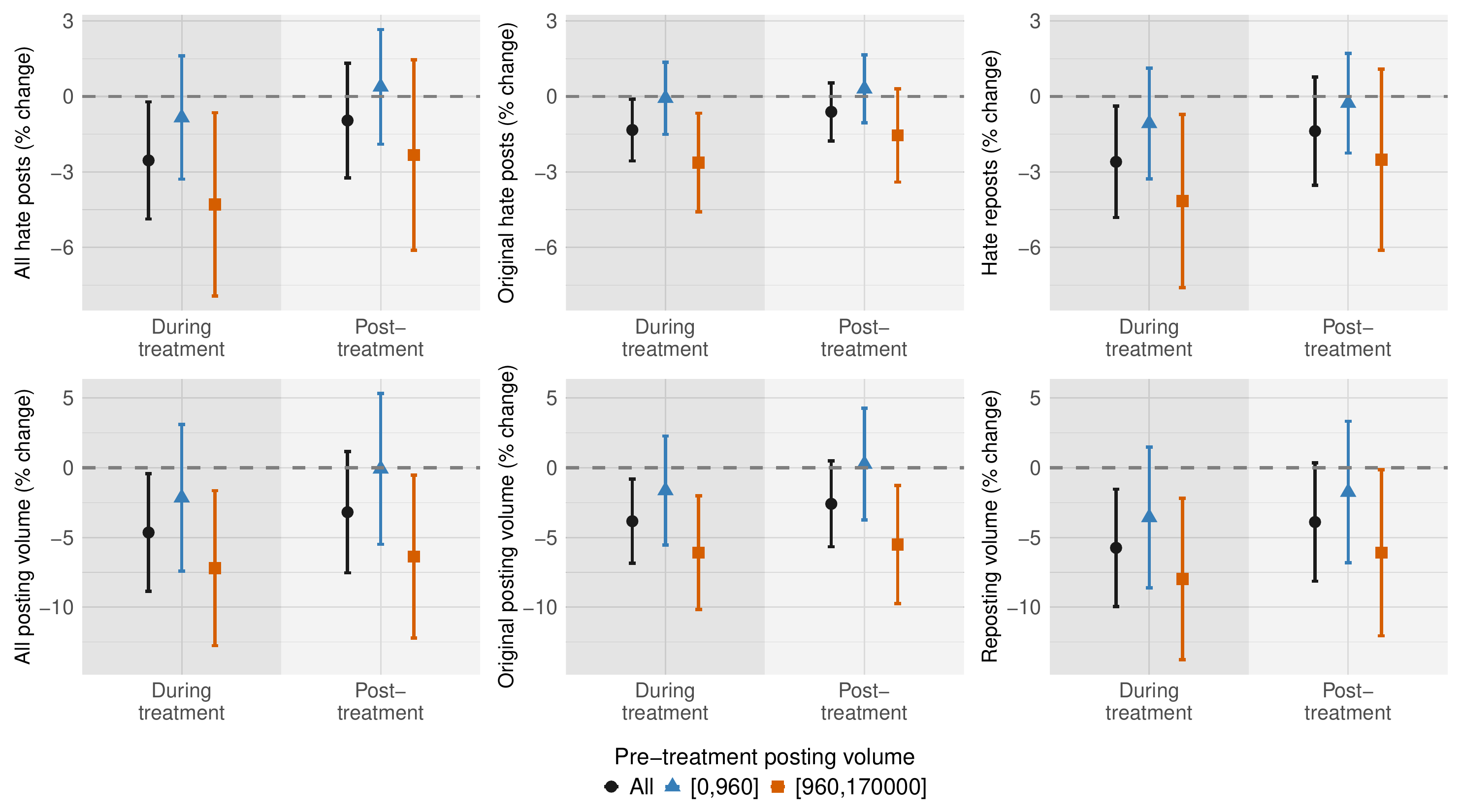}
    \caption{Heterogeneity in treatment effect on hate outcomes (top row) and posting volume (bottom row) by pre-treatment activity. Each panel reports estimated effects for high- and low-activity users (above/below the median) across different outcomes. Point estimates are expressed as percentage change and bars correspond to 95\% confidence intervals.}
    \label{fig:heterogeneity_total_activity}
\end{figure}

\section{Heterogeneity of direct effects}
\label{sec:si_heterogeneity_direct}

In the main text, we document heterogeneity in direct treatment effects on hate posts. In this section, we extend the heterogeneity analysis to the remaining outcomes analyzed in the main text: total posting volume, original posts, reposts, original hate posts, and hate reposts. Specifically, we examine how treatment effects vary across two key dimensions of user behavior before the intervention: (i) overall posting volume and (ii) pre-treatment hate share. We divide users into those above and below the median of each heterogeneity variable. Treatment effects within each subgroup are estimated using the same framework as described in the main text and section \ref{sec:si_direct_effect}. This analysis assesses whether more active users or users with relatively lower engagement in hate content respond differently to the campaign across the full set of outcome measures.

\subsection{Heterogeneity by pre-treatment activity}

\newcolumntype{Y}{>{\centering\arraybackslash}X}

\begin{table}[t]
  \centering
  \small
  \setlength{\tabcolsep}{6pt}
  \renewcommand{\arraystretch}{1.5}
  \caption{Average treatment effects (ATE), standard errors (SE), and \emph{p}-values by pre-treatment activity for hate posts and posting volume outcomes during the treatment period. Estimates are in log units and are expressed as percentage change. Each cell lists ATE on the first line and (SE, \emph{p}-value) on the second.}
  \label{tab:heterogeneity_activity}

  \begin{tabularx}{\textwidth}{l YYY}
    \toprule
    \textbf{Outcome} & \textbf{All} & \textbf{Low} & \textbf{High} \\
    \midrule

    Total hate &
      \makecell{-2.54\\(1.19, $p=0.032$)} &
      \makecell{-0.84\\(1.24, $p=0.50$)} &
      \makecell{-4.29\\(1.86, $p=0.021$)} \\
    
    Original hate &
      \makecell{-1.33 \\(0.627, $p=0.033$)} &
      \makecell{-0.070 \\(0.731, $p=0.92$)} &
      \makecell{-2.62 \\(1.00, $p=0.009$)} \\ 

    Reposted hate &
      \makecell{-2.60 \\(1.13, $p=0.021$)} &
      \makecell{-1.07\\(1.12, $p=0.34$)} &
      \makecell{-4.16 \\(1.76, $p=0.018$)} \\

      \midrule
      \midrule

    Total posts &
      \makecell{-4.65\\(2.15, $p=0.030$)} &
      \makecell{-2.15\\(2.68, $p=0.42$)} &
      \makecell{-7.20\\(2.83, $p=0.011$)} \\
    
    Original posts &
      \makecell{-3.84\\(1.54, $p=0.013$)} &
      \makecell{-1.64\\(1.99, $p=0.41$)} &
      \makecell{-6.09\\(2.08, $p=0.003$)} \\
     
     Reposts &
      \makecell{-5.75\\(2.15, $p=0.007$)} &
      \makecell{-3.57\\(2.57, $p=0.16$)} &
      \makecell{-7.98\\(2.95, $p=0.007$)} \\
    \bottomrule
  \end{tabularx}
\end{table}

Fig.~\ref{fig:heterogeneity_total_activity} and table \ref{tab:heterogeneity_activity} present treatment effects for users grouped by pre-treatment posting activity. Across all  outcomes, treatment effects are directionally consistent with those reported in the main text: users with above-median activity levels exhibit larger, statistically significant reductions in hate content and total posting, whereas effects among less active users are smaller and not statistically significant. These patterns indicate that the intervention was more effective among highly engaged users, who were both more exposed to the ads and more likely to generate measurable changes in their online behavior. Although differences between high- and low-activity groups are not statistically significant when directly compared, the point estimates consistently show stronger responses among active users. This reinforces the interpretation that engagement intensity amplifies responsiveness to prosocial messaging.

\subsection{Heterogeneity by pre-treatment hate share}

\begin{figure}[t]
    \centering
    \includegraphics[width=\textwidth]{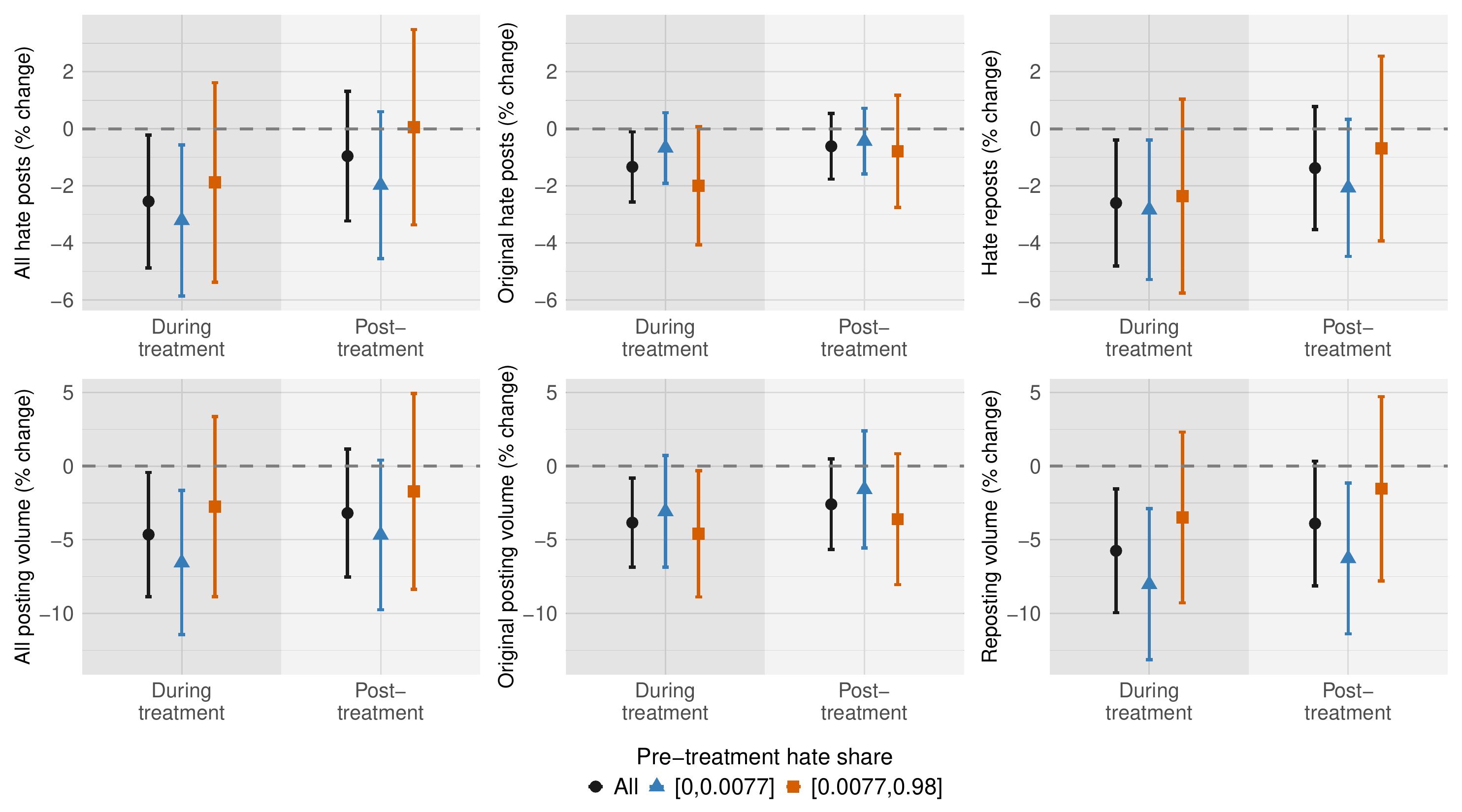}
    \caption{Heterogeneity of treatment effect on hate outcomes (top row) and posting volume (bottom row) by pre-treatment hate share. Each panel reports estimated effects for users above and below the median share of hate content in pre-treatment activity. Point estimates are expressed as percentage change and bars correspond to 95\% confidence intervals.}
    \label{fig:heterogeneity_hate_fraction}
\end{figure}

Fig.~\ref{fig:heterogeneity_hate_fraction} and table \ref{tab:heterogeneity_hatefraction} report analogous results when users are grouped by their pre-treatment hate share. The effects again mirror those in the main text: the campaign reduced hate content more effectively among users whose baseline activity contained a smaller share of hate. For users with hate fractions above the median, effects are weaker and generally insignificant, consistent with these users being less responsive to normative messages. However, less hate-prone users show measurable reductions in both hateful and overall posting behavior.

\begin{table}[t]
  \centering
  \small
  \setlength{\tabcolsep}{6pt}
  \renewcommand{\arraystretch}{1.5}
  \caption{Average treatment effects (ATE), standard errors (SE), and \emph{p}-values by pre-treatment hate share for all outcomes during the treatment period. Estimates are in log units and are expressed as percentage change. Each cell lists ATE on the first line and (SE, \emph{p}-value) on the second.}
  \label{tab:heterogeneity_hatefraction}

  \begin{tabularx}{\textwidth}{l YYY}
    \toprule
    \textbf{Outcome} & \textbf{All} & \textbf{Low} & \textbf{High} \\
    \midrule

    Total hate &
      \makecell{-2.54\\(1.19, $p=0.032$)} &
      \makecell{-3.21\\(1.35, $p=0.017$)} &
      \makecell{-1.88\\(1.78, $p=0.29$)} \\

    Original hate &
      \makecell{-1.33 \\(0.627, $p=0.033$)} &
      \makecell{-0.67\\(0.630, $p=0.29$)} &
      \makecell{-1.99 \\(1.06,  $p=0.059$)} \\

    Reposted hate &
      \makecell{-2.60\\(1.13, $p=0.021$)} &
      \makecell{-2.84\\(1.25, $p=0.023$)} &
      \makecell{-2.36\\(1.73, $p=0.17$)} \\

    \midrule
    \midrule
    
    Total posts &
      \makecell{-4.65\\(2.15, $p=0.030$)} &
      \makecell{-6.55\\(2.49, $p=0.009$)} &
      \makecell{-2.75\\(3.12, $p=0.37$)} \\

    Original posts &
      \makecell{-3.84\\(1.54, $p=0.013$)} &
      \makecell{-3.08\\(1.93, $p=0.11$)} &
      \makecell{-4.60\\(2.18, $p=0.035$)} \\

    Reposts &
      \makecell{-5.75\\(2.15, $p=0.007$)} &
      \makecell{-8.02\\(2.61, $p=0.002$)} &
      \makecell{-3.48\\(2.96, $p=0.24$)} \\

    \bottomrule
  \end{tabularx}
\end{table}

\section{Indirect effects among participants}
\label{sec:si_indirect_participants}

This section describes how we define and estimate indirect effects among participants (Fig. \ref{fig:heterogeneity_direct}C). We first introduce the causal estimand and the role of exposure mappings, then describe the weighted $q$-fraction exposure model used to operationalize exposure conditions, define the treated-low, control-high, and treated-high exposure effects as formal estimands, explain how exposure propensities were obtained, and, finally, summarize the estimation procedure based on the weighted regression approach of \cite{gao2025causal}. Throughout, we assume local interference, meaning that a user’s outcome may depend on their own assignment and on the assignments of their direct neighbors, but not on units beyond one hop.
 
\subsection{Estimand and exposure mapping}

Let $Z=(Z_1,\ldots,Z_N)$ denote the random treatment assignment vector and let $Y_i(z)$ denote the potential outcome of participant $i$ under assignment vector $z$. The target estimand compares the hypothetical worlds in which all users are treated, $z=(1, 1, \dots, 1)$, versus all are untreated, $z=(0, 0, \dots, 0)$,
\begin{align*}
\tau = \frac{1}{N}\sum_{i=1}^N \left[ Y_i(1,1,\ldots,1) - Y_i(0,0,\ldots,0)\right].
\end{align*}
The mapping $z\mapsto Y_i(z)$ is high-dimensional and cannot be estimated directly since both worlds cannot be observed at the same time. To address this, we often use an exposure mapping, a function $T_i=T(i,Z,A)$ that maps the full assignment vector and the network $A$ into a low-dimensional exposure category $t$ for each user $i$. Under a correctly specified mapping, the potential outcome on the full treatment vector $z$ reduces to $Y_i(t)$, meaning that only the exposure category matters for the user's outcome. This reduction allows us to define causal estimands as differences in mean potential outcomes across exposure categories rather than across full assignment vectors. More importantly, if the exposure mapping function is relatively simple, e.g., under a local interference assumption, all exposure conditions will be observable within a single randomization schedule and thus identifiable.

\subsection{Weighted $q$-fraction exposure model}

To classify exposure conditions, we adopt a weighted $q$-fraction exposure model with $q=0.7$, which generalizes the $q$-fraction framework of \citet{ugander2013graph} to weighted networks. For each user $i$, let $W_i$ denote the sum of edge weights to all of $i$'s neighbors in the interaction network as described in section \ref{si:network_construction}. Let $W_{i,T}$ denote the total weight towards treated neighbors. User $i$ is assumed to be exposed to treatment indirectly when at least 70\% of its neighbors, weighted by their edge weights, receive the treatment, i.e. $W_{i,T}/W_i \ge 0.7$. Using this rule, a user can be assigned to one of four exposure conditions.
\begin{itemize}
\item \textbf{Control-low exposure:} The user did not receive the treatment, and at least 70\% of their neighbors were also untreated.
\item \textbf{Control-high exposure:} The user did not receive the treatment, but at least 70\% of their neighbors were treated.
\item \textbf{Treated-low exposure:} The user received the treatment, but at least 70\% of their neighbors were untreated.
\item \textbf{Treated-high exposure:} The user received the treatment, and at least 70\% of their neighbors were treated.
\end{itemize}

This weighted $q$-fraction rule approximates the idea that outcomes depend on whether a sufficiently large share of influential neighbors receive treatment, rather than requiring that all neighbors receive treatment. 

\subsection{Exposure contrasts}

Let $\mu(t)=\frac{1}{N}\sum_{i=1}^n Y_i(t)$ denote the mean potential outcome under exposure condition $t$. The treated-low, control-high, and treated-high exposure effects are defined as \emph{contrasts} in these mean potential outcomes. The treated-low (TL) effect compares treated and untreated users when their neighbors are effectively untreated,
\begin{align}
\label{eq:si_direct_contrast}
\tau_{\mathrm{TL}} = \mu(\text{treated-low exposure}) - \mu(\text{control-low exposure})
\end{align}

The control-high (CH) effect compares untreated users with treated neighbors to untreated users with untreated neighbors,
\begin{align}
\label{eq:si_indirect_contrast}
\tau_{\mathrm{CH}} = \mu(\text{control-high exposure}) - \mu(\text{control-low exposure})
\end{align}

The treated-high (TH) effect compares treated users with treated neighbors to untreated users with untreated neighbors,
\begin{align}
\label{eq:si_full_contrast}
\tau_{\mathrm{TH}} = \mu(\text{treated-high exposure}) - \mu(\text{control-low exposure})
\end{align}

Because exposure depends on both individual and neighbor treatments, the random neighborhood variation created by the hole punching cluster randomization is necessary to identify all three estimands.

\subsection{Propensity scores}

For each user $i$ and exposure category $t$, let $\pi_i(t)=\Pr(T_i=t)$ denote the probability that user $i$ falls into exposure $t$ under the hole punching randomization design. These probabilities are known from the design but cannot be computed analytically because of the two-stage assignment and the weighted $q$-fraction exposure condition. We therefore estimate $\pi_i(t)$ by running 50,000 Monte Carlo simulations of the randomization procedure. For each simulation, we redraw the treatment at the cluster level, independently flip each assignment with probability 0.18, recompute exposure categories for all users, and record the frequency with which each user attains each exposure. The estimated propensity is the empirical fraction of simulations in which $T_i=t$. We also derive an analytical method based on dynamic programming, similar to the one described in \citet{ugander2013graph} for the q-fraction exposure model. However, this dynamic programming is more complex due to the hole punching step and the weighted exposure rule and is therefore not included here. Nevertheless, we confirmed that the propensities estimated via Monte Carlo closely match those obtained analytically. The estimated $\pi_i(t)$ determine inverse-propensity weights $w_i = 1/\pi_i(T_i)$ for estimation.

\subsection{Estimation via weighted regression}
To estimate the target estimand $\tau(t)$ for each exposure category in equations \ref{eq:si_direct_contrast}, \ref{eq:si_indirect_contrast}, and \ref{eq:si_full_contrast}, we apply the weighted regression estimator of \citet{gao2025causal} with inverse propensities as weights, which becomes the Hájek estimator in our setting. Let $z_i(t)$ be an indicator that user $i$ is observed in exposure $t$. Weighted least squares is performed by regressing $Y_i$ on the full set of exposure indicators, using weights $w_i = 1/\pi_i(T_i)$. 

In our implementation, this regression uses the same difference-in-differences outcome described earlier and includes dummy variables for each exposure as exogenous variables, with 40 equal-sized binned pre-treatment posting volume as a covariate to improve precision.
Under this estimator, the fitted coefficient corresponding to exposure $t$ is equal to the Hájek inverse-probability weighted mean,
\begin{align*}
\hat{\mu}(t) = 
\frac{\sum_{i} z_i(t) w_i Y_i}{\sum_{i} z_i(t) w_i}.
\end{align*}

This estimation approach has two advantages. First, it is equivalent to the Hájek estimator specified in our preregistration, while providing a unified regression-based framework that can incorporate covariates if desired. Second, as \citet{gao2025causal} shows, the estimator retains desirable design-based properties under interference, and its variance can be consistently estimated using the network-robust HAC variance estimator. 

Given the estimated average potential outcomes under each exposure, we obtain the contrasts corresponding to the treated-high, control-high, and treated-low exposure effects by taking simple differences of the relevant $\hat{\mu}(t)$ values. We compute the standard errors for these contrasts using the same network-robust HAC covariance matrix estimator from the weighted regression, assuming that contrasts of exposure means are linear combinations of the regression coefficients.

Figure~\ref{fig:heterogeneity_direct} in the main text shows our results. Specifically, we found no statistically significant indirect effect ($\tau_{\text{CH}}$) under this exposure model. The results remain qualitatively unchanged when using an alternative choice, $q=0.8$ fraction, of exposure threshold.

\subsection{Turnover in reposted users}
\label{sec:si_turnover}

Because the analysis spans a long period from January 1, 2023 to August 31, 2024, it is reasonable to expect that interaction patterns will evolve over the course of the study. We therefore explore how rapidly the set of accounts reposted by participants changes over time. For each participant $u$ and month $t$, let $\mathcal{S}_u(t)$ denote the set of accounts reposted by $u$ in month $t$. We define the renewal rate as
\begin{equation}
    r_u(t) = 1 - \frac{\lvert \mathcal{S}_u(t) \cap \mathcal{S}_u(t+1) \rvert}{\lvert \mathcal{S}_u(t) \rvert},
    \label{eq:user_renewal}
\end{equation}
which captures the share of $u$'s reposted accounts in month $t$ that do not appear again in month $t+1$. To test whether participants have a relatively stable core set of accounts they repost, we also compute renewal conditioned on the top $q$ percentile of accounts with whom $u$ interacts most frequently. That is, we condition $\mathcal{S}_u(t)$ on the set of accounts accounting for the top $q$ interactions during month $t$.

\begin{figure}[t]
    \centering
	\includegraphics[width=0.7\textwidth]{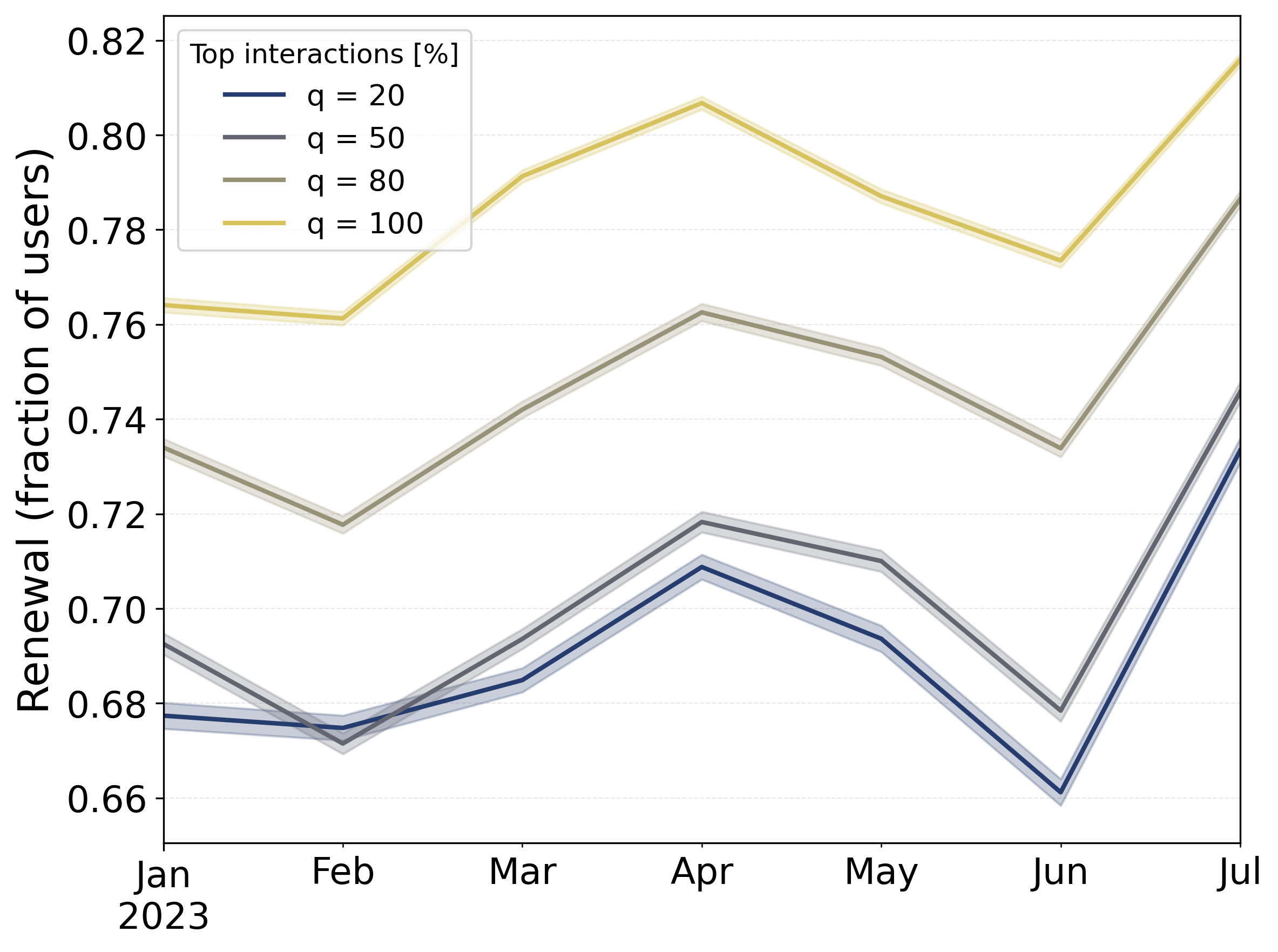}
    \caption{Monthly renewal of accounts retweeted by participants. Renewal is measured as the fraction of accounts retweeted in month $t$ that are not retweeted in month $t+1$, for different conditioning levels based on the most frequently interacted accounts (see legend). The shaded area reports the 95\% confidence intervals.}
	\label{fig:rct_churn_monthly}
\end{figure}

Fig.~\ref{fig:rct_churn_monthly} shows the average renewal rate across participants for different conditioning levels. Without conditioning, the average monthly renewal rate is 0.78, indicating that a large share of reposted accounts changes from one month to the next. Renewal remains high even when restricting attention to the most engaged accounts: when focusing on the top 20\% of interactions, the average renewal rate is 0.68. These results indicate that the set of accounts reposted by participants changes substantially over time, implying that interaction networks constructed at a single point in time provide only a transient snapshot. The magnitude of these renewal rates closely aligns with the high weekly user renewal documented by \cite{kolic2025chambers}.

Moreover, only 3.39\% of the accounts reposted by participants were themselves participants, indicating that most relevant interaction pathways lay outside the experimental sample.
The high turnover rate combined with the low rate at which participants share each other's content provides a plausible explanation for why we do not detect indirect effects among participants.

% The second perspective concerns upstream users; i.e., the top 400 non-participants whose content was frequently reposted by participants in the pre-treatment period defined as in Section \label{sec:si_indirect_upstream}. For each user in each month, we compute the fraction of their reposters, here referred to as \emph{audience}, who are participants. Decreases in this fraction over time would suggest that participants are gradually disengaging from the upstream user and allocating attention to other accounts.

% Fig.~\ref{fig:producers_rct_churn_monthly} shows the evolution of the average fraction of audience for the selected 400 upstream users. On average, approximately half 51\% of the audience consists of users during the pre-treatment period. This fraction decreased to 42\% during the treatment period and further decreased to 38\% during the post-treatment period. This pattern is consistent with the turnover observed in participants' reposted accounts and suggests that participants steadily replace the upstream users they engage with during the study period.

% \begin{figure}[t]
%     \centering
% 	\includegraphics[width=0.7\textwidth]{images/si/producers_churn_month_to_month.png}
%     \caption{\textbf{Average fraction of upstream users' audience composed of participants over time.} The decreasing trend in this fraction indicates increasing disengagement of participants from previously retweeted producers. The shaded area reports the 95-CI.}
% 	\label{fig:producers_rct_churn_monthly}
% \end{figure}

\section{Indirect effects on upstream users}
\label{sec:si_indirect_upstream}

\subsection{Selection of upstream users}

To identify candidate upstream users, we collected all hate content reposted by the 80{,}154 participants during the pre-treatment period from August 1 to November 28, 2023. From this collection, we extracted the original authors and excluded participants. Then, for the remaining authors, we identified all users who had reposted any of their original content during the same period. To avoid including extremely high-activity accounts for whom participants constitute only a negligible share of total engagement, we restricted attention to users with fewer than 15{,}000 posts in the pre-treatment period. For each remaining user, we computed the \emph{repost-share fraction}, defined as the share of all pre-treatment reposts they received that were attributed to participants. Formally, let $\mathcal{O}_i$ be the set of original posts created by upstream user $i$ pre-treatment, and for each original post $k \in \mathcal{O}_i$, let $\mathcal{R}_{ik}$ denote the set of all users who reposted that post. The repost-share fraction is
\begin{align}
F_i \;=\;
\frac{\sum_{k \in \mathcal{O}_i} \;\sum_{u \in \mathcal{R}_{ik}} \mathbf{1}\{u \in \mathrm{participants}\} }
     {\sum_{k \in \mathcal{O}_i} | \mathcal{R}_{ik}| } 
     \label{eq:si_producer_potential_exposure}
\end{align}
We define the primary analysis set as the 400 upstream users with the highest $F_i$ values. Because $F_i$ reflects the maximum possible fraction of a user's audience that is exposed to the experiment, it also represents the user's maximal potential exposure to treatment, which we analyze in the next section.

%An alternative preregistered definition, based on the absolute number of RCT reposts an upstream user received, did not yield substantial effects (SI Section~XXXX), likely because high absolute volume does not translate into a large \emph{share} of attention from RCT users.

\subsection{Exposure definition}

For each upstream user $i$, we define the exposure to the experiment analogously to the repost-share fraction $F_i$, but now incorporating the treatment status of each participant reposter. In other words, exposure is the share of pre-treatment reposts attributable to \emph{treated} participants:
\begin{align}
T_i \;=\;
\frac{\sum_{k \in \mathcal{O}_i} \;\sum_{u \in \mathcal{R}_{ik}} 
      \mathbf{1}\{u \in \mathrm{participants}\}\,\mathbf{1}\{u \text{ treated}\} }
     {\sum_{k \in \mathcal{O}_i} \;\sum_{u \in \mathcal{R}_{ik}} 
      \mathbf{1}\{u \in \mathrm{participants}\}}.
      \label{eq:si_producer_exposure}
\end{align}
Thus, $T_i$ denotes the fraction of an upstream user's pre-treatment engagement attributable to users assigned to the treatment. Higher values of $T_i$ indicate greater potential indirect exposure for user $i$, since a larger share of their pre-treatment audience consisted of treated participants. Note that $0 \le T_i \le F_i \le 1$ for any upstream user $i$.

\subsection{Test statistic for randomization inference}

For each outcome variable, we estimate the marginal association between exposure $T_i$ and the upstream user's outcome using a regression-based test statistic. To address direct effects and improve statistical power, we employ a difference estimator whenever a valid pre-treatment measure is available. Let $Y_i^{\text{during}}$ denote the outcome during the treatment period (estimation for post-treatment period is similar) and $Y_i^{\text{pre}}$ the corresponding pre-treatment value. For outcomes where such a pre-treatment baseline is defined (e.g., log hate reposts by participants or non-participants), we form the adjusted outcome
\begin{align}
\Delta_i^{\text{during}} \;=\; Y_i^{\text{during}} - Y_i^{\text{pre}}
\end{align}
and estimate the test statistic using
\begin{align}
\Delta_i^{\text{during}} \;=\; \alpha \;+\; \theta\, T_i \;+\; \gamma_{s(i)} \;+\; \varepsilon_i,
\label{eq:upstream_diff_reg}
\end{align}
where $\gamma_{s(i)}$ are fixed effects for bins of pre-treatment total posting volume by the upstream user. This covariate adjustment improves precision by removing level differences across users that are unrelated to treatment exposure.

For outcomes without a meaningful pre-treatment baseline, such as the count of lost reposters, which is defined relative to the pre-treatment audience, we use the raw outcome $Y_i^{\text{during}}$ and estimate:
\begin{align}
Y_i^{\text{during}} \;=\; \alpha \;+\; \theta\, T_i \;+\; \gamma_{s(i)} \;+\; \varepsilon_i
\label{eq:upstream_raw_reg}
\end{align}

In all cases, we use the coefficient $\widehat{\theta}$ as the \emph{test statistic} for randomization inference.
It is not itself the causal estimand, but rather the statistic whose distribution is evaluated under the randomization procedure.

\subsection{Causal estimand}

We assume a linear exposure–response model
\begin{align}
Y_i(T_i) \;=\; Y_i(0) \;+\; \tau\, T_i,
\end{align}
where $\tau$ represents the marginal change in the outcome for a one-unit increase in exposure (i.e., all pre-treatment reposters receive the treatment). Because $T_i$ ranges from 0 to 1, results are reported per one-percentage-point increase in $T_i$. A sharp null hypothesis $H_0: \tau = \tau_0$ fully specifies all potential outcomes.

\subsection{Randomization inference}

To test the sharp null hypothesis $H_0 : \tau = \tau_0$ for any given value of $\tau_0$ (including $\tau_0=0$ for no indirect effects) and to construct confidence intervals for $\tau$, we implement randomization inference (RI) as follows:

\begin{enumerate}
\item Replicate the two‑stage assignment with hole punching 10{,}000 times, recomputing $T_i^{(b)}$ for each simulated assignment $b$ and each upstream user $i$, via equation~\eqref{eq:si_producer_exposure}.
\item For each assignment $b$, compute adjusted outcomes $\widetilde{\Delta}^{(b)}$ that would have observed if the null outcome model were true.
For each $\tau_0$, compute  
\begin{align*}
\widetilde{\Delta}_i^{(b)}(\tau_0) \;=\; \Delta_i - \tau_0 T_i + \tau_0 T_i^{(b)}
\end{align*}
where $T_i$ is the exposure of upstream user $i$ under the actual treatment assignment, and $T_i^{(b)}$ is their simulated exposure under the new assignment $b$.
\item For each $b$, regress $\widetilde{\Delta}^{(b)}(\tau_0)$ on $T_i^{(b)}$ with the same activity-bin fixed effects $\gamma_{s(i)}$, and record the slope estimates $\widehat{\theta}^{(b)}(\tau_0)$.
\item The empirical distribution of $\{\widehat{\theta}^{(b)}(\tau_0)\}_{b=1}^{10{,}000}$ approximates the distribution of the test statistic under sharp null $H_0$.
\end{enumerate}

The RI $p$‑value is the two‑sided tail probability of the observed $\widehat{\theta}$ under the null distribution $\{\widehat{\theta}^{(b)}\}_{b=1}^{10{,}000}$ \citep{rosenbaum2007interference}. By testing $H_0:\tau=\tau_0$ over a grid of $\tau_0$ and inverting the test (i.e., finding the acceptance region), we obtain confidence intervals for the causal parameter $\tau$.
This approach ensures exact inference by generating the null distribution under the dependence structure induced by cluster randomization and hole punching. Furthermore, it explicitly accounts for potential indirect effects and requires no modeling assumptions beyond the sharp null and the linear exposure-response outcome model. All RI results reported in Fig.~\ref{fig:producers_heterogeneity} and in the Supplementary Information follows this procedure.

\section{Persistence of indirect effects}
\label{sec:si_persistence_indirect}

As for the direct effect, we define a persistence parameter as the fraction of the indirect effect on upstream users during treatment that persists into the post-treatment period. Since our indirect effect analysis relies on a randomization-inference (RI) procedure, we adapt the procedure of \citet{lin2025persuading} to the RI setting and estimate confidence intervals for the persistence parameter by inverting the RI test. Below, we assume that for each upstream user $i$, we observe outcomes during and after the intervention, denoted $Y_i^{\text{during}}$ and $Y_i^{\text{post}}$, such as log hate reposts or log loss of pre-treatment reposters.

\subsection{Conceptual approach}
\label{sec:si_persistence_method_indirect_effect}
As in Section~\ref{sec:si_indirect_upstream}, let $T_i$ denote upstream user $i$'s exposure (the share of $i$'s audience that is treated under the two-stage design with hole punching). We assume a linear exposure--response model in which the indirect effect during the campaign is $\tau$ and the post-treatment indirect effect is a fraction $\beta \in [0,1]$ of the during-treatment effect:
\begin{align*}
Y_i^{\text{during}}(T_i) &= Y_i^{\text{during}}(0) + \tau T_i \\
Y_i^{\text{post}}(T_i,\beta) &= Y_i^{\text{post}}(0) + \beta \tau T_i
\end{align*}
Here, $Y_i^{\text{during}}(0)$ and $Y_i^{\text{post}}(0)$ are the potential outcomes of upstream user $i$ under zero exposure during and post-treatment. The persistence parameter $\beta$ captures the fraction of the effect during treatment that persists after the campaign.

\subsubsection{Conditioning on $\tau$.} The randomization inference (RI) for $\beta$ requires constructing, for each simulated re-randomization of treatment assignment, the corresponding exposure $T_i$ and the implied potential outcomes in both periods. Under the model above, these potential outcomes depend on the during-treatment indirect effect size $\tau$. Without specifying $\tau$, we cannot impute the counterfactual outcomes $Y_i^{\text{during}}(T_i)$ and $Y_i^{\text{post}}(T_i,\beta)$ under alternative assignments, and therefore cannot generate the RI distribution for the persistence statistic.

For this reason, we conduct RI over a grid of plausible $\tau$ values. For each candidate $\tau$, we (i) impute during- and post-treatment outcomes under each simulated assignment for a sequence of $\beta_0$ values and (ii) invert the resulting RI tests to obtain a confidence interval for $\beta$ conditional on that $\tau$. We then report, for each assumed $\tau$, the estimated $\beta$, its RI confidence interval, and the RI $p$-value for the null of no persistence.

\subsubsection{Pre-treatment adjustment and differenced outcomes.}
As throughout the paper, we increase precision by working with changes relative to a pre-treatment baseline. Let $Y_i^{\text{pre}}$ denote upstream user $i$'s pre-period outcome, and define
\begin{align*}
\Delta_i^{\text{during}} &= Y_i^{\text{during}} - Y_i^{\text{pre}} \\
\Delta_i^{\text{post}} &= Y_i^{\text{post}} - Y_i^{\text{pre}}
\end{align*}

\subsubsection{RI test statistic}
The test statistic used in RI procedure is the coefficient on $\Delta_i^{\text{during}}$ in the following regression model:
\begin{align}
\Delta_i^{\text{post}} = \alpha + \theta\,\Delta_i^{\text{during}} + \gamma_{s(i)} + \varepsilon_i
\label{eq:si_indirect_persistence_test_stat}
\end{align}
where $\gamma_{s(i)}$ is the same pre-treatment activity-bin fixed effects used in the indirect-effect analysis. We can simulate the randomization procedure, for each randomization impute the outcomes given the value of $\beta_0$ under the sharp null, and estimate the test statistic. This allows us to construct the sharp null distribution of $\theta$ test statistic and compare it against the observed value.

\begin{table}[h!]
\centering
\caption{RI-based persistence estimates for indirect effects by assumed during-treatment effect $\tau$. For each outcome and $\tau$, we report the 95\% RI confidence interval and the $p$-value for $H_0:\beta=0$. Randomization inference is conducted over $0 \le \beta \le 1$, hence the upper bound of 1.}
\label{tab:si_indirect_persistence_by_tau}
\begin{tabular}{clccc}
\toprule
Assumed $\tau$ & Outcome & 95\% CI & $p$-value \\
\midrule
\multicolumn{4}{l}{\textbf{Panel A: Log hate reposts}} \\
\midrule
-0.01 & Log hate reposts (participants)     & (0, 1) & 0.55  \\
-0.01 & Log hate reposts (non-participants) & (0, 1) & 0.29  \\
\addlinespace
-0.02 & Log hate reposts (participants)     & (0, 1) &  0.11 \\
-0.02 & Log hate reposts (non-participants) & (0.09, 1) &  0.039 \\
\addlinespace
-0.03 & Log hate reposts (participants)     & (0.29, 1)  &  0.0097 \\
-0.03 & Log hate reposts (non-participants) & (0.39, 1) &  0.0061 \\
\addlinespace
-0.04 & Log hate reposts (participants)     & (0.43, 1) &  0.0007 \\
-0.04 & Log hate reposts (non-participants) & (0.5, 1) & 0.0003 \\
\addlinespace
-0.05 & Log hate reposts (participants)     & (0.5, 0.99) &  $<10^{-6}$ \\
-0.05 & Log hate reposts (non-participants) & (0.55, 1) & $<10^{-6}$ \\
\midrule\midrule
\multicolumn{4}{l}{\textbf{Panel B: Log hate reposters lost}} \\
\midrule
0.03 & Log hate reposters lost (participants)     & (0, 1) & 0.75 \\
0.03 & Log hate reposters lost (non-participants) & (0, 1) & 0.35 \\
\addlinespace
0.04 & Log hate reposters lost (participants)     & (0, 1) & 0.11 \\
0.04 & Log hate reposters lost (non-participants) & (0.91, 1) & 0.008 \\
\addlinespace
0.05 & Log hate reposters lost (participants)     & (0.92, 1) & 0.0065 \\
0.05 & Log hate reposters lost (non-participants) & (0.96, 1) & 0.0001 \\
\addlinespace
0.06 & Log hate reposters lost (participants)     & (0.97, 1) & 0.0001 \\
0.06 & Log hate reposters lost (non-participants) & (0.98, 1) & $< 10^{-6}$ \\
\addlinespace
0.07 & Log hate reposters lost (participants)     & (0.98, 1) & $< 10^{-6}$ \\
0.07 & Log hate reposters lost (non-participants) & (0.99, 1) & $< 10^{-6}$ \\
\bottomrule
\end{tabular}
\end{table}

\subsubsection{Randomization inference}
We implement RI for the sharp null hypothesis $H_0:\beta=\beta_0$ conditional on a candidate during-treatment effect $\tau$. The implementation mirrors the RI procedure for the indirect effect in Section~\ref{sec:si_indirect_upstream}: for each simulated re-randomization we (i) recompute exposures and (ii) construct the outcomes that would have been observed under the null outcome model by ``shifting'' the observed outcomes from the realized exposure $T_i$ to the simulated exposure $T_i^{(b)}$.
For a fixed pair $(\tau,\beta_0)$, proceed as follows:

\begin{enumerate}
\item Replicate the original two-stage assignment with hole punching $B=10{,}000$ times. For each simulated assignment $b$, compute upstream user exposures $T_i^{(b)}$ via equation~\eqref{eq:si_producer_exposure}.
\item For each $b$, construct the adjusted differenced outcomes that would have been observed if $\tau$ and $\beta_0$ were true.
Specifically, define
\begin{align*}
\widetilde{\Delta}_i^{\text{during},(b)}(\tau) 
&= \Delta_i^{\text{during}} - \tau T_i + \tau T_i^{(b)} \\
\widetilde{\Delta}_i^{\text{post},(b)}(\tau,\beta_0) 
&= \Delta_i^{\text{post}} - \beta_0 \tau T_i + \beta_0 \tau T_i^{(b)}
\end{align*}
where $T_i$ is upstream user $i$'s exposure under the realized assignment and $T_i^{(b)}$ is upstream user $i$'s exposure under the simulated assignment $b$.
\item For each $b$, estimate the persistence regression
\begin{align*}
\widetilde{\Delta}_i^{\text{post},(b)}(\tau,\beta_0)
= \alpha^{(b)} + \widehat{\theta}^{(b)}(\tau,\beta_0)\,
\widetilde{\Delta}_i^{\text{during},(b)}(\tau)
+ \gamma_{s(i)} + \varepsilon_i^{(b)}
\end{align*}
including the same activity-bin fixed effects $\gamma_{s(i)}$ used in the indirect-effect analysis, and record the slope $\widehat{\theta}^{(b)}(\tau,\beta_0)$ as the RI test statistic.
\item The empirical distribution of $\{\widehat{\theta}^{(b)}(\tau,\beta_0)\}_{b=1}^{B}$ approximates the null distribution of the test statistic under $H_0:\beta=\beta_0$ conditional on $\tau$.
\item The randomization $p$-value for $H_0 : \beta = \beta_0$ is given by
\begin{align*}
p(\beta_0) = \frac{1}{B} \sum_{b=1}^B
\mathbf{1}\bigl\{ \bigl|\hat\theta^{(b)}(\tau, \beta_0)\bigr| \ge \bigl|\hat\theta(\tau, \beta_0)\bigr| \bigr\}
\end{align*}
which compares the observed test statistic $\hat\theta(\tau, \beta_0)$ to its null distribution.
\item By repeating Steps 1–4 over a grid of candidate values $\beta_0$ and collecting all values for which $H_0 : \beta = \beta_0$ is not rejected at level $\alpha$ (e.g., $\alpha = 0.05$), we obtain a $(1-\alpha)$ confidence interval for the persistence parameter $\beta$ conditional on indirect effect $\tau$.
\end{enumerate}

\subsection{Results}
We applied the randomization-inference procedure described above to estimate the persistence parameter $\beta$ for each upstream user outcome across a grid of plausible $\tau$ values, as reported in the main text. Table~\ref{tab:si_indirect_persistence_by_tau} reports the 95\% confidence intervals obtained by inverting the RI tests and the p-value associated with the sharp null of no persistence. Across all outcomes and almost all $\tau$ values, we find substantial persistence of the indirect effects into the post-treatment period, indicating that the reduction in downstream amplification continues after the intervention ends.

\begin{figure*}[t!]
\centering
\begin{subfigure}[t]{\textwidth}
\centering
\includegraphics[width=\textwidth]{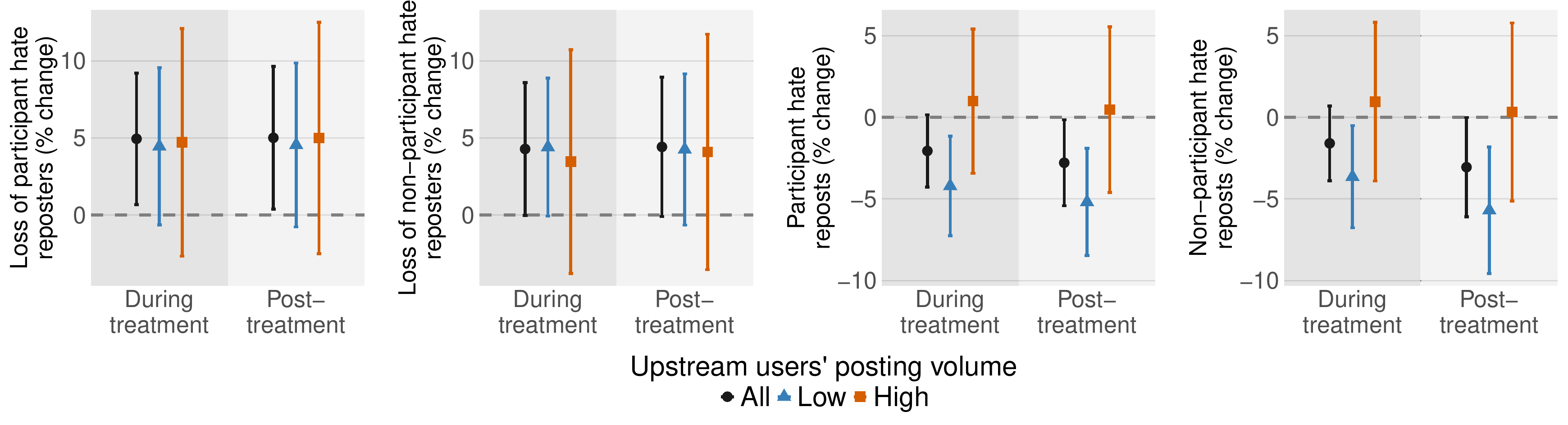}
\end{subfigure}

\vspace*{0.5em}

\begin{subfigure}[t]{\textwidth}
\centering
\includegraphics[width=\textwidth]{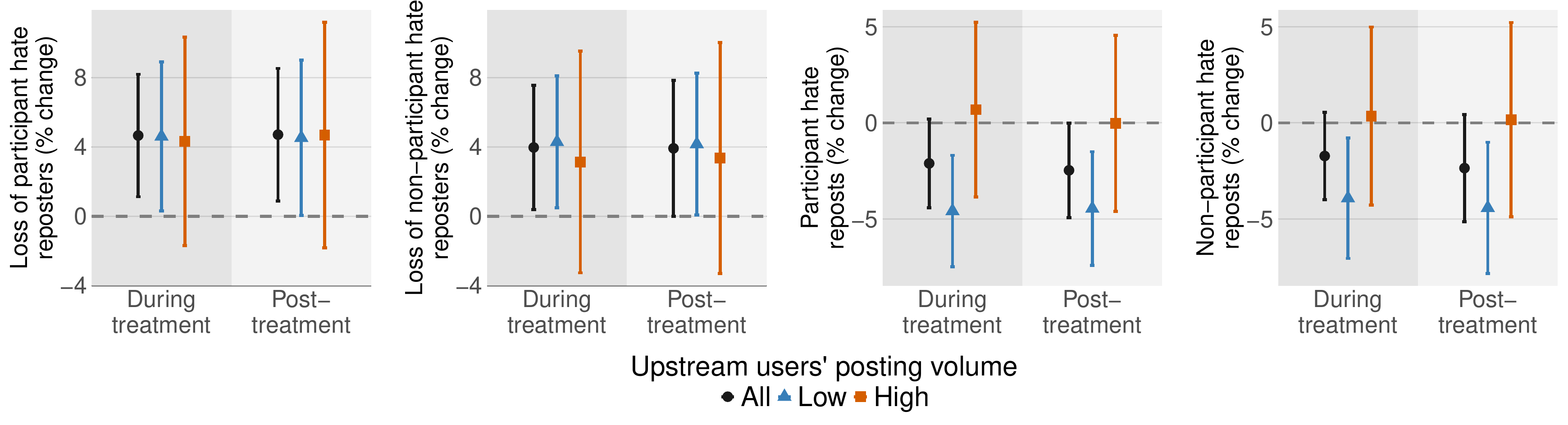}
\end{subfigure}

\vspace*{0.5em}

\begin{subfigure}[t]{\textwidth}
\centering
\includegraphics[width=\textwidth]{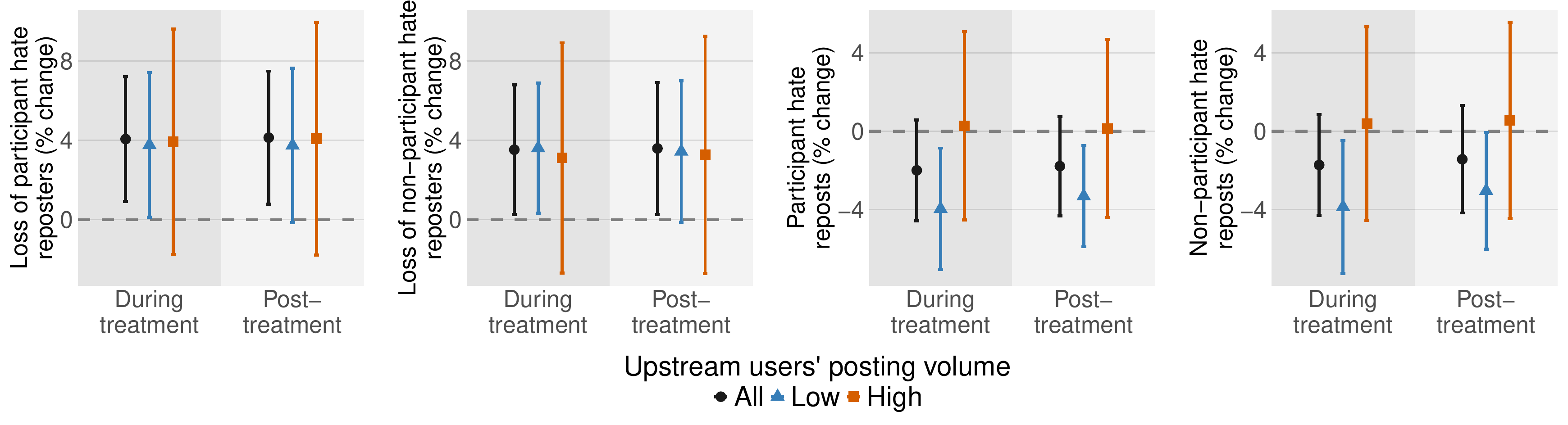}
\end{subfigure}
\caption{Indirect effects on upstream users under alternative binary hate thresholds 0.1 (top), 0.75 (middle), and 0.9 (bottom). Panels report percentage changes associated with a one-percentage-point increase in the share of an upstream user's pre-treatment audience that was treated. Outcome definitions are the same as in Figure \ref{fig:producers_heterogeneity} of the main text. Error bars show 95\% RI confidence intervals.}
\label{fig:si_indirect_robust_thresholds}
\end{figure*}

\section{Robustness of indirect effects}
\label{sec:si_robustness_indirect_upstream}

This section investigates the robustness of our indirect effect estimates on upstream users to a series of alternative measurement and design choices. Throughout, we follow the same two-stage randomization with hole punching and the same RI-based inference procedure as in Section~\ref{sec:si_indirect_upstream}, using the same activity-bin fixed effects $\gamma_{s(i)}$. We report effects as percentage changes associated with a one-percentage-point increase in an upstream user's treated audience share, consistent with Figure \ref{fig:producers_heterogeneity} in the main text.

\subsection{Alternative hate measures}
The main text presents reposts of an upstream user's original content using a hate score threshold of 0.5. We assess robustness using: (i) alternative binary thresholds and (ii) a continuous, score-weighted outcome.

\paragraph{Binary thresholds.} We redefine an upstream user's hate content using different binary thresholds. Specifically, for each original post by an upstream user, we assign a hate indicator equal to 1 if the model's hate score exceeds the threshold $c \in \{0.1, 0.75, 0.90\}$, and 0 otherwise. For each threshold, we recompute hate reposts received on original hate posts and repeat the same RI-based estimation for indirect effects on the four outcomes of interest (Figure~\ref{fig:si_indirect_robust_thresholds}). Across thresholds, the estimated indirect effects remain comparable in sign and magnitude to those in the main text.

\begin{figure}[t]
\centering
\includegraphics[width=0.6\linewidth]{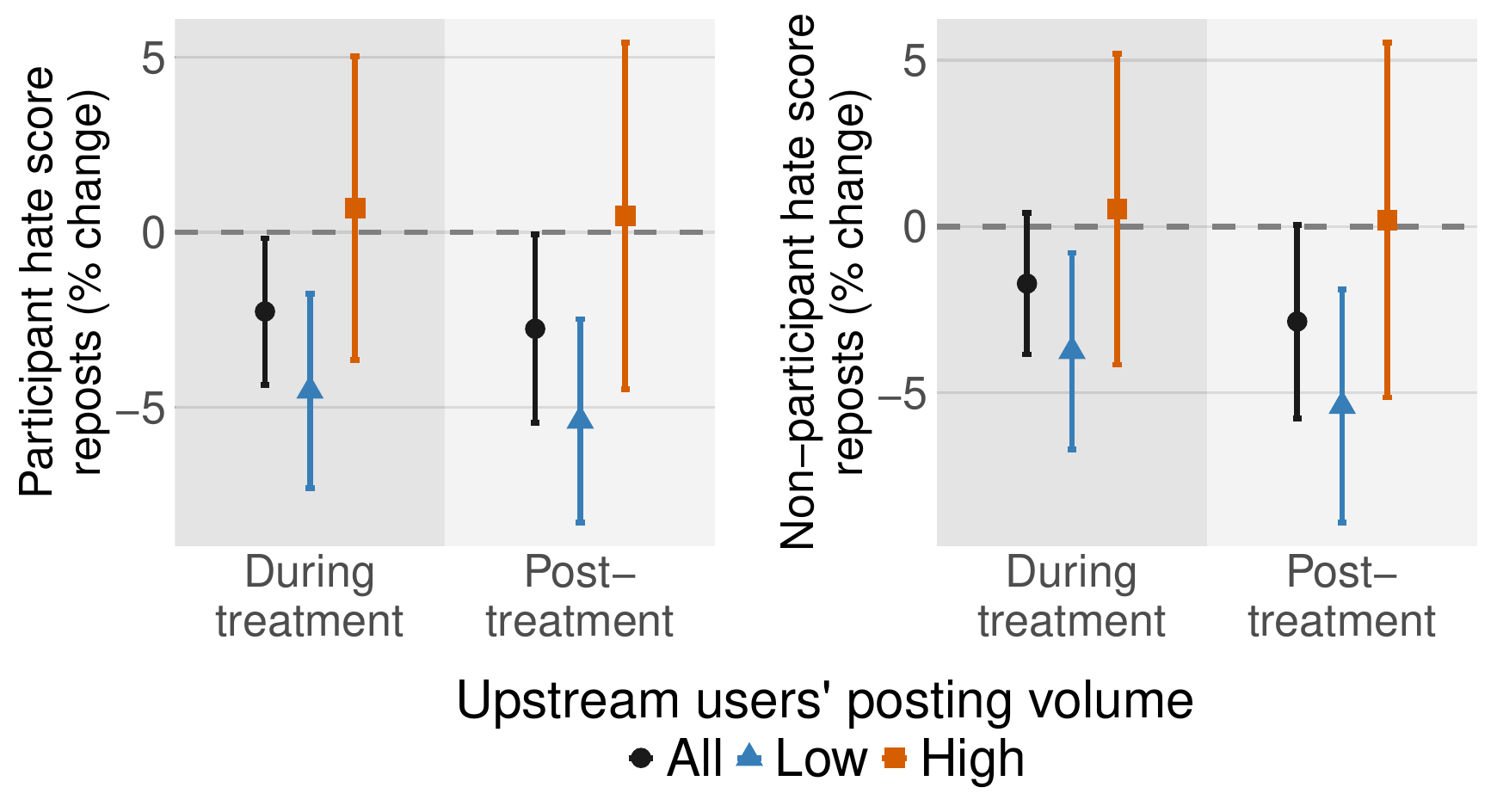}
\caption{Indirect effects using a continuous hate-score outcome.
Panels report percentage changes associated with a one-percentage-point increase in treated audience share, where the outcome is the sum of hate scores over reposted content received by each upstream user. Error bars show 95\% RI confidence intervals.}
\label{fig:si_indirect_robust_hatescore}
\end{figure}

\paragraph{Continuous hate-score outcome.} We also replace the binary hate repost count with a continuous, score-weighted measure. Specifically, instead of counting only reposts of posts with score above a threshold, we sum the hate scores of all reposted original posts received by an upstream user, yielding an outcome that up-weights more hateful reposts while retaining information below any fixed cutoff. Figure~\ref{fig:si_indirect_robust_hatescore} reports RI-based estimates for participants and non-participants using this score-weighted outcome. The estimated effect on hate reposts is a decline of 2.26\% for participants ($p = 0.037$) and 1.72\% for non-participants ($p = 0.098$), consistent with the main estimates.

\subsection{Alternative sample selection}
\begin{figure*}[t]
\centering
\begin{subfigure}[t]{\textwidth}
\centering
\includegraphics[width=\textwidth]{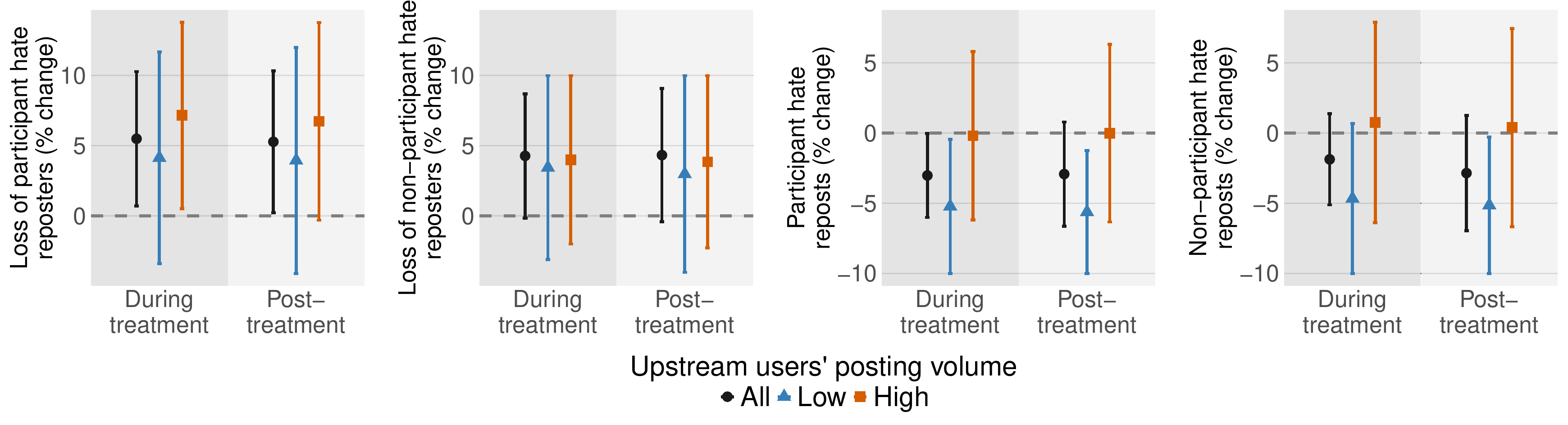}
\caption{Top 200 upstream users by exposure}
\end{subfigure}

\vspace*{2em}

\begin{subfigure}[t]{\textwidth}
\centering
\includegraphics[width=\textwidth]{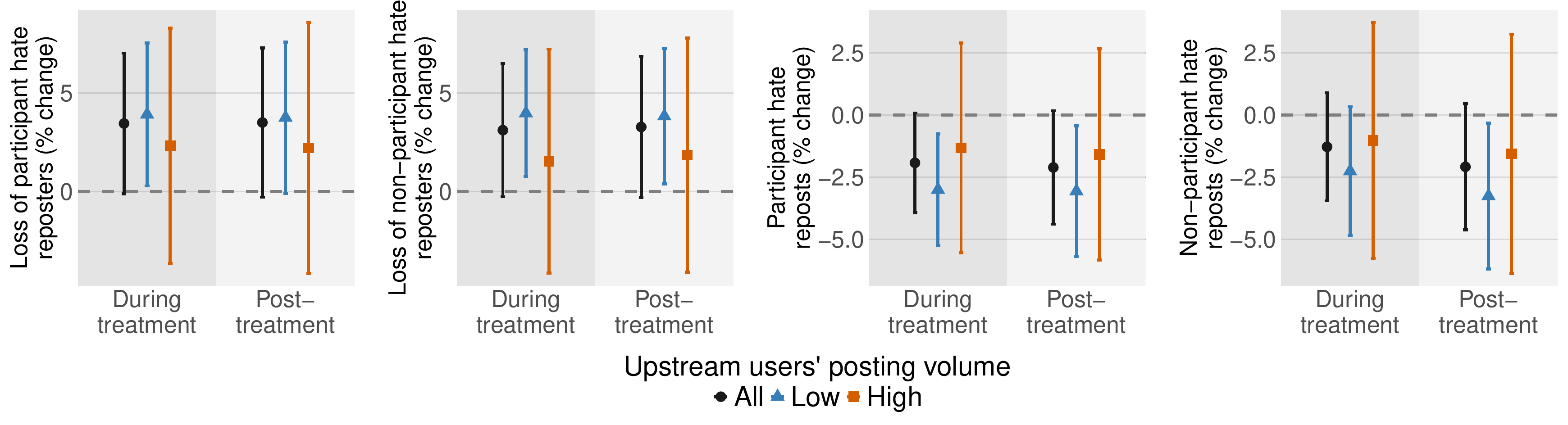}
\caption{Top 600 upstream users by exposure}
\end{subfigure}
\caption{Indirect effects on top 200 (top) and 600 (bottom) upstream users with highest potential exposure to the treatment. Panels report percentage changes associated with a one-percentage-point increase in the share of an upstream user's pre-treatment audience that was treated. Outcome types are defined in Figure \ref{fig:producers_heterogeneity}. Error bars show 95\% RI confidence intervals.}
\label{fig:si_indirect_robust_top200_600}
\end{figure*}

The preregistered analysis presented in the main text focuses on the top 400 upstream users ranked by potential exposure to the experiment participants. To assess robustness to this choice, we repeat the analysis using alternative samples: the top 200 and the top 600 upstream users by potential exposure measure defined in equation \ref{eq:si_producer_potential_exposure}. Figure~\ref{fig:si_indirect_robust_top200_600} shows the corresponding estimates. The results closely track the findings in Figure \ref{fig:producers_heterogeneity} in the main text. For example, among the top 200 users, hate reposts by participants and non-participants declined by 3\% ($p=0.047$) and 1.9\% ($p=0.25$) respectively, and the loss of pre-treatment hate reposters among participants and non-participants increased by 5.5\% ($p=0.034$) and 4.3\% ($p=0.055$), respectively per one-percentage-point increase in treated audience share.

\subsection{Alternative exposure definition}
\begin{figure*}[t!]
\centering
\includegraphics[width=\textwidth]{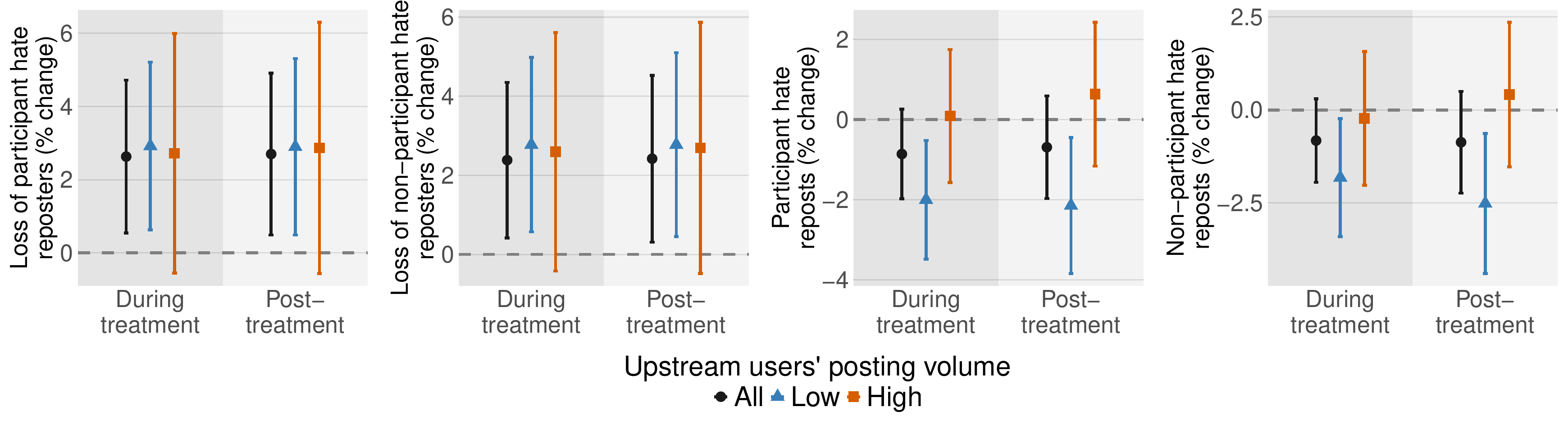}
\caption{Indirect effects using a hate-weighted exposure definition.
Upstream users are selected as the top 400 by the alternative measure in equation~\eqref{eq:si_producer_potential_exposure_hate}, and exposure to treatment is defined by equation~\eqref{eq:si_producer_exposure_hate}. Error bars show 95\% RI confidence intervals.}
\label{fig:si_indirect_robust_exposure}
\end{figure*}

In the main text and according to our preregistration, upstream users were selected based on their potential exposure to participants, measured as the share of their pre-treatment reposts attributable to participants. A natural alternative is to weigh each repost by the hate score of the original content, so that engagement with more hateful content contributes more to measured exposure. We therefore modify our original measure of potential exposure in equation \ref{eq:si_producer_potential_exposure} and 
define an alternative hate-weighted measure:
\begin{align}
F_i \;=\;
\frac{\sum_{k \in \mathcal{O}_i} \;\sum_{u \in \mathcal{R}_{ik}} \mathbf{1}\{u \in \mathrm{RCT}\}\, h_{ik}}
     {\sum_{k \in \mathcal{O}_i} \;\sum_{u \in \mathcal{R}_{ik}} h_{ik}}
\label{eq:si_producer_potential_exposure_hate}
\end{align}
where $h_{ik}$ is the hate score for original post $k$ by upstream user $i$, $\mathcal{O}_i$ is the set of $i$'s pre-treatment original posts, and $\mathcal{R}_{ik}$ is the set of users who reposted post $k$ during the pre-treatment period. Analogously, we modify the realized exposure in equation~\ref{eq:si_producer_exposure} to define a hate-weighted share of experiment participant repost activity attributable to \emph{treated} users
\begin{align}
T_i \;=\;
\frac{\sum_{k \in \mathcal{O}_i} \;\sum_{u \in \mathcal{R}_{ik}}
      \mathbf{1}\{u \in \mathrm{RCT}\}\,\mathbf{1}\{u \text{ treated}\}\, h_{ik}}
     {\sum_{k \in \mathcal{O}_i} \;\sum_{u \in \mathcal{R}_{ik}}
      \mathbf{1}\{u \in \mathrm{RCT}\}\, h_{ik}}
\label{eq:si_producer_exposure_hate}
\end{align}
Using this alternative potential exposure $F_i$, we selected the top 400 upstream users and re-ran the RI-based indirect effect analysis exactly as in Section~\ref{sec:si_indirect_upstream}, replacing the exposure measure with the hate-weighted definition in equation~\ref{eq:si_producer_exposure_hate}. Figure~\ref{fig:si_indirect_robust_exposure} reports results under this alternative exposure definition. For an additional one-percentage-point increase in treated audience according to this exposure measure, hate reposts declined by 0.86\% ($p = 0.14$) among participants and by 0.83\% ($p = 0.16$) among non-participants. The loss of pre-treatment hate reposters increases by 2.7\% ($p = 0.014$) for participants and by 2.4\% ($p = 0.019$) for non-participants. As in the main text, heterogeneity by upstream posting volume persists: effects are concentrated among lower-activity users, while high-activity users exhibit statistically insignificant declines.

\begin{table}[t]
\centering
\caption{Robustness of indirect effects to the choice of pre-treatment period.
Entries report the 95\% RI confidence interval for the outcome during the treatment period; parentheses report the two-sided RI $p$-value for the sharp null of zero effect.}
\label{tab:si_indirect_robust_pretreat}
\begin{tabular}{lcccc}
\toprule
 & \multicolumn{2}{c}{Loss of hate reposters} & \multicolumn{2}{c}{Hate reposts} \\
\cmidrule(lr){2-3} \cmidrule(lr){4-5}
Pre-treatment window &
Participants &
Non-participants &
Participants &
Non-participants \\
\midrule

\makecell{2023-06-01 to\\2023-11-28} &
\makecell{$[-1.0, 11.1]\%$  \\ ($p=0.078$)} &
\makecell{$[-2.0, 10.0]\%$  \\ ($p=0.12$)} &
\makecell{$[-5.7, -0.28]\%$ \\ ($p=0.037$)} &
\makecell{$[-5.0, 0.54]\%$  \\ ($p=0.097$)} 
\vspace{6pt} \\

\makecell{2023-07-01 to\\2023-11-28} &
\makecell{$[-0.64, 10]\%$   \\ ($p=0.065$)} &
\makecell{$[-1.1, 9.3]\%$   \\ ($p=0.095$)} &
\makecell{$[-6.5, -0.55]\%$ \\ ($p=0.023$)} &
\makecell{$[-6.1, 0.03]\%$  \\ ($p=0.051$)} 
\vspace{6pt} \\

\makecell{2023-08-01 to\\2023-11-28 \\ (main text)} &
\makecell{$[1.6, 9.2]\%$ \\ ($p=0.013$)} &
\makecell{$[0.68, 8.3]\%$ \\ ($p=0.025$)} &
\makecell{$[-5.4, -0.54]\%$ \\ ($p=0.019$)} &
\makecell{$[-4.4, 0.24]\%$ \\ ($p=0.075$)} \\

\bottomrule
\end{tabular}
\end{table}

\subsection{Alternative pre-treatment period}
Because audiences exhibit substantial churn, as shown in section \ref{sec:si_turnover}), pre-treatment measures should be as close as possible to the intervention if they are to predict post-treatment reposting behavior. For this reason, our primary specification uses the pre-treatment window August 1 to November 28, 2023 to form pre-treatment adjustments. In this section, we re-estimate indirect effects using two alternative pre-treatment windows that go further back in time: (i) July 1, 2023 to November 28, 2023, and (ii) June 1, 2023 to November 28, 2023. Table~\ref{tab:si_indirect_robust_pretreat} reports RI-based estimates for the four upstream outcomes studied in the main text: change in hate reposts by experiment participants and non-participants, loss of pre-treatment reposters among participants and non-participants during the treatment period. Each cell reports the RI-based 95\% RI confidence interval along with the two-sided RI $p$-value for the sharp null of zero effect. The final row corresponds to the main-text specification.
Across alternative pre-treatment windows, the estimated indirect effects are similar in magnitude and direction.

\begin{figure}[t]
\centering
\includegraphics[width=0.3\linewidth]{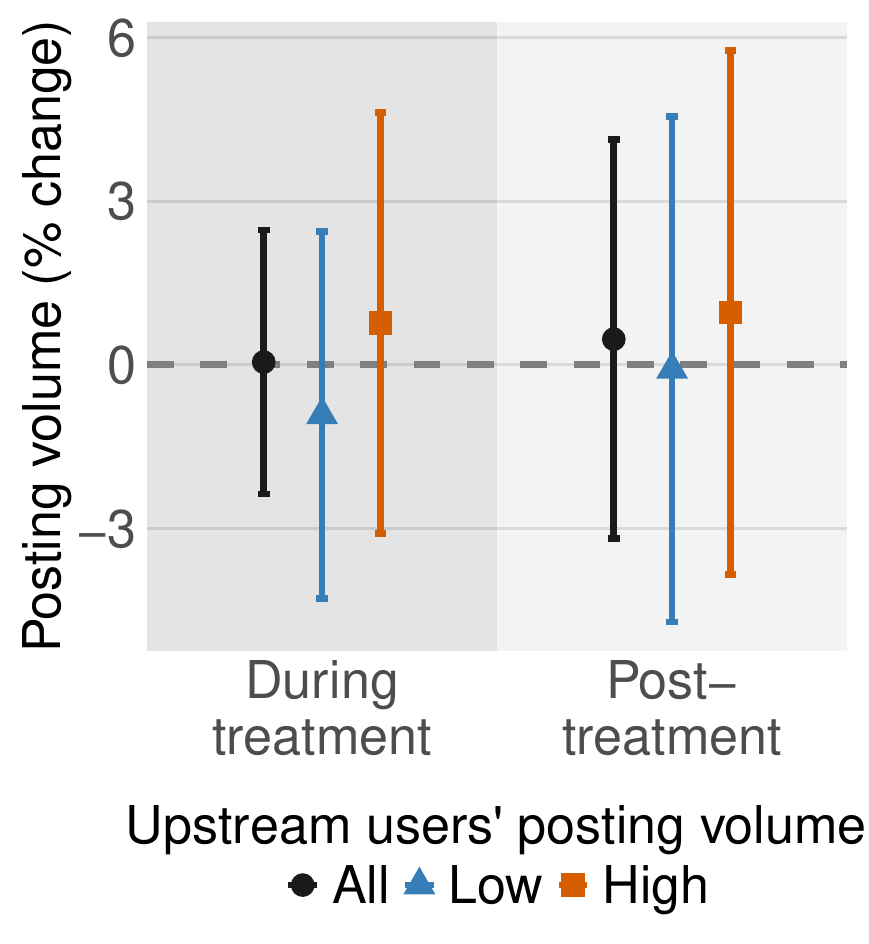}
\caption{Negative control test using upstream users' total posting volume.
The figure reports the marginal percentage change associated with a one-percentage-point increase in treated audience share. Heterogeneity is with respect to pre-treatment posting volume. Error bars show 95\% RI confidence intervals.}
\label{fig:si_indirect_robust_placebo}
\end{figure}

\subsection{Negative control test}
Finally, we implemented two placebo tests for which we did not expect the intervention to have an effect. First, we repeated the RI analysis using overall posting activity of upstream users as a placebo outcome.
Since the campaign ads only targeted experimental participants and never reached upstream users directly, we did not expect any changes in their posting volume. Figure~\ref{fig:si_indirect_robust_placebo} reports the RI-based estimates for this placebo outcome. Consistent with expectations, estimated effects are small and statistically indistinguishable from zero ($p=0.75$).

As a second placebo test, we estimated indirect effects during periods that entirely precede the intervention. We applied the same RI-based difference estimation procedure described above, and computed placebo effects over two alternative period specifications:

\begin{enumerate}
    \item A placebo specification with a pre-treatment adjustment period from 2023-01-01 to 2023-04-01, a placebo during-treatment period from 2023-04-01 to 2023-08-01, and a placebo post-treatment period from 2023-08-01 to 2023-11-01.
    \item A placebo specification with a pre-treatment adjustment period from 2023-04-01 to 2023-07-01 and a placebo during-treatment period from 2023-07-01 to 2023-11-01. In this case, the available time span does not allow for a separate placebo post-treatment period.
\end{enumerate}

\begin{figure}[t]
    \centering
    \begin{subfigure}[t]{\linewidth}
        \centering
        \includegraphics[width=\linewidth]{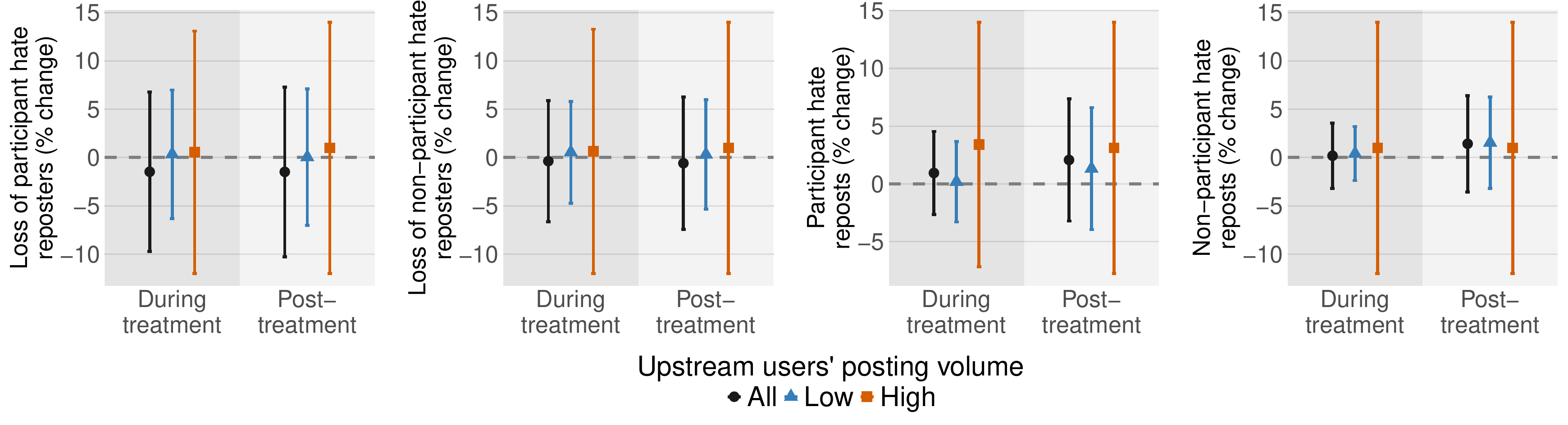}
        \label{fig:si_placebo_spec1}
    \end{subfigure}
    \hfill
    \begin{subfigure}[t]{\linewidth}
        \centering
        \includegraphics[width=\linewidth]{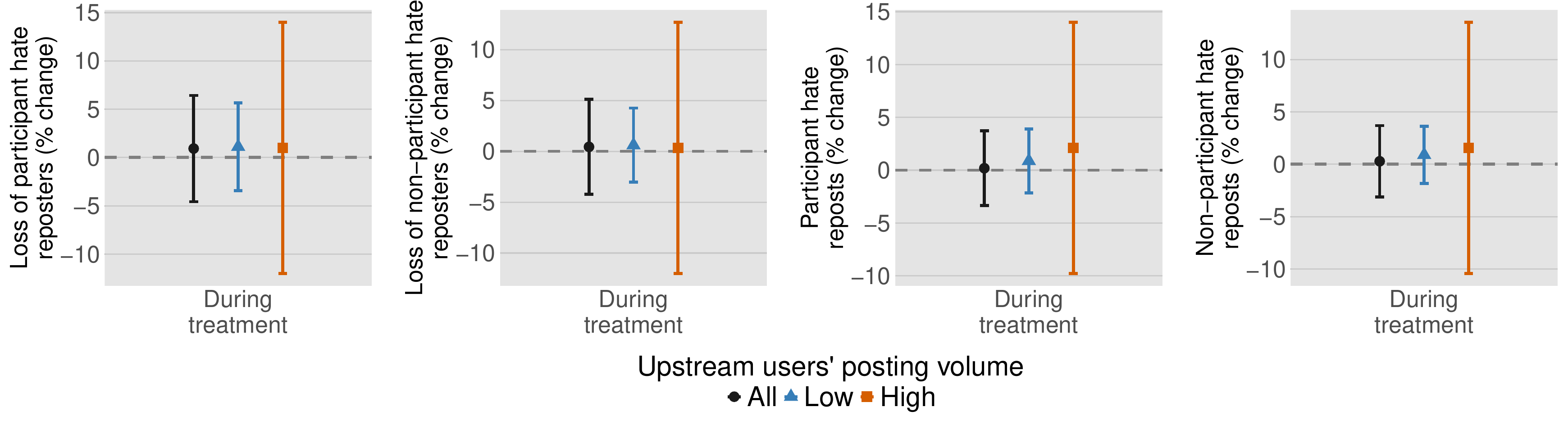}
        \label{fig:si_placebo_spec2}
    \end{subfigure}

    \caption{Placebo tests using pre-intervention periods.
Panels report RI-based indirect effect estimates computed over periods that precede the experiment launch. The top panel corresponds to specification 1, with pre-treatment from 2023-01-01 to 2023-04-01 which includes both placebo during- and placebo post-treatment periods. The bottom panel corresponds to specification 2, with pre-treatment from 2023-04-01 to 2023-07-01 which includes only a placebo during-treatment period. Error bars show 95\% RI confidence intervals.}
    \label{fig:si_indirect_robust_placebo_pre}
\end{figure}

Because treatment had not yet occurred during any of these periods, we expected no systematic relationship between upstream exposure and outcomes in either specification. Figure~\ref{fig:si_indirect_robust_placebo_pre} reports the RI-based estimates for these placebo periods. Consistent with expectation, estimated effects for both participant and non-participant outcomes are close to zero and statistically indistinguishable from zero across hate reposts and loss of hate reposters. Table~\ref{tab:si_indirect_placebo_pre} reports the corresponding point estimates and RI $p$-values.

\begin{table}[t]
\centering
\caption{RI-based placebo estimates using pre-intervention periods. For comparison, the last row includes the estimates from the main text using the correct period specification.
Entries report the 95\% RI confidence interval for the outcome during the treatment placebo period; parentheses report the two-sided RI $p$-value for the sharp null of zero effect.}
\label{tab:si_indirect_placebo_pre}
\begin{tabular}{lcccc}
\toprule
 & \multicolumn{2}{c}{Loss of hate reposters} & \multicolumn{2}{c}{Hate reposts} \\
\cmidrule(lr){2-3} \cmidrule(lr){4-5}
Pre-treatment window &
Participants &
Non-participants &
Participants &
Non-participants \\
\midrule
\midrule

\multicolumn{5}{l}{\emph{Specification 1: }} \\
\addlinespace

\makecell{2023-01-01 to\\2023-04-01} &
\makecell{$[-9.72, 6.76]\%$  \\ ($p=1$)} &
\makecell{$[-6.64, 5.89]\%$  \\ ($p=0.80$)} &
\makecell{$[-2.66, 4.54]\%$  \\ ($p=0.68$)} &
\makecell{$[-3.21, 3.57]\%$  \\ ($p=0.99$)} 
\vspace{6pt} \\

\midrule
\multicolumn{5}{l}{\emph{Specification 2: }} \\
\addlinespace

\makecell{2023-04-01 to\\2023-07-01} &
\makecell{$[-4.58, 6.43]\%$  \\ ($p=0.56$)} &
\makecell{$[-4.23, 5.11]\%$  \\ ($p=0.73$)} &
\makecell{$[-3.36, 3.73]\%$  \\ ($p=0.89$)} &
\makecell{$[-3.12, 3.68]\%$  \\ ($p=0.75$)} 

\vspace{6pt} \\

\midrule
\multicolumn{5}{l}{\emph{Main text specification: }} \\
\addlinespace

\makecell{2023-08-01 to\\2023-11-28} &
\makecell{$[1.6, 9.2]\%$ \\ ($p=0.013$)} &
\makecell{$[0.68, 8.3]\%$ \\ ($p=0.025$)} &
\makecell{$[-5.4, -0.54]\%$ \\ ($p=0.019$)} &
\makecell{$[-4.4, 0.24]\%$ \\ ($p=0.075$)} \\

\bottomrule
\end{tabular}
\end{table}

\section{Testing for repost chain interruption}
\label{sec:si_rank_reposte}

The main text documents indirect effects on upstream users that extend to \emph{non-participants} who were never targeted by the campaign. We consider two competing mechanisms for this non-participant effect. First, this could arise mechanically from the interruption of repost cascades: if treated participants repost less, some downstream non-participants may simply never encounter the upstream post through the usual repost chain. Second, treated participants may reduce their reposting in a way that the platform interprets as a decline in popularity of upstream users' content, reducing how often that content is shown and thereby lowering reposts overall, including among non-participants. Because the second mechanism involves platform internals, it is difficult to test directly without access to ranking and distribution systems. Here, we investigate implications of the first mechanism and find no evidence consistent with repost chain disruption, which increases the plausibility of a platform response explanation.

The test for repost chain interruption rests on the following intuition: if treated participants were breaking repost chains, then some non-participants who would have reposted after being exposed to participant reposts would no longer do so. In that case, the composition of non-participant reposters who remain shifts more toward users who repost independently (e.g., by encountering the original upstream users' posts directly), rather than through exposure to participant reposts. As a result, upstream users with higher exposure to the treatment should exhibit a shift in the relative ordering of participant and non-participant reposters, with non-participants reposting less often \emph{after} participants.

We operationalize this idea by constructing, for each upstream user, a measure of how often non-participant reposts occur after participant reposts. Fix an upstream user $i$ and let $\mathcal{O}_i$ denote the set of original posts by $i$. For each original post $k \in \mathcal{O}_i$, let $\mathcal{P}_{ik}$ be the set of participant users who reposted $k$, $\mathcal{N}_{ik}$ the set of non-participant users who reposted $k$, and let $t_{iku}$ denote the time of user $u$'s repost of $k$. We define the ordering outcome as

\begin{align*}
R_i
=
\frac{
\sum_{k \in \mathcal{O}_i}
\;\sum_{u \in \mathcal{N}_{ik}}
\;\sum_{v \in \mathcal{P}_{ik}}
\mathbf{1}\!\left\{ t_{ikv} < t_{iku} \right\}
}{
\sum_{k \in \mathcal{O}_i} \left|\mathcal{N}_{ik}\right| \left|\mathcal{P}_{ik}\right|
}
\end{align*}
This quantity has a simple probabilistic interpretation: for a randomly selected original post $k$, it is the probability that a randomly selected participant repost occurs before a randomly selected non-participant repost of the same post. Values closer to one indicate that non-participants tend to repost \emph{after} participants (more consistent with repost cascades flowing through participants), whereas values closer to zero indicate that non-participants tend to repost \emph{before} participants.

\begin{figure}[t]
\centering
\includegraphics[width=0.3\linewidth]{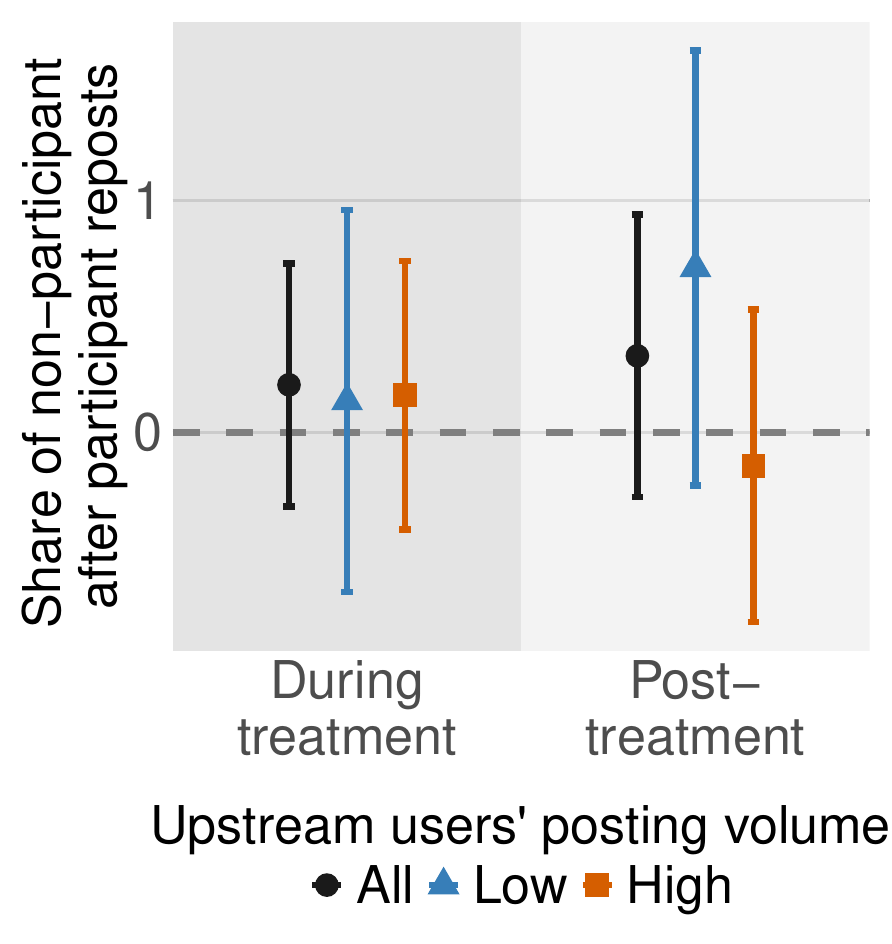}
\caption{Indirect effects on the relative ordering of participant and non-participant reposts. The figure reports the marginal percentage change associated with a one-percentage-point increase in treated audience share. Heterogeneity is with respect to pre-treatment posting volume. Error bars show 95\% RI confidence intervals.}
\label{fig:si_repost_chain}
\end{figure}

We compute $R_i$ as an upstream-user-level outcome and test whether higher exposure causes changes in the relative ordering of participant versus non-participant reposters using the same randomization-inference methodology with pre-treatment difference adjustment described in Section~\ref{sec:si_indirect_upstream}. Figure~\ref{fig:si_repost_chain} shows that there is no statistically significant effect in either the during- or post-treatment periods (95\% confidence interval of $[-0.32, 0.73]$ with $p=0.54$ during treatment period, and 95\% confidence interval of $[-0.28, 0.94]$ with $p=0.34$ post-treatment).
These null results are inconsistent with the repost chain interruption account, and instead support the interpretation that reduced participant engagement triggers a platform response that lowers visibility of the upstream user.

\section{Cost-effectiveness}
\label{sec:si_cost_effectiveness}

This section provides an illustrative comparison between the cost-effectiveness of our intervention and conventional platform moderation. The intent is not to estimate moderation costs precisely, which are not publicly observable, but to benchmark the intervention against the approximate scale and documented effectiveness of platform moderation efforts.

We approximate global moderation costs using publicly reported figures indicating that X employed roughly 2,500 human moderators in early 2023. Assuming a conservative wage of USD 10 per hour, 8-hour workdays, and continuous daily coverage, this implies a approximate global moderation budget of USD 6 million per month. Despite these investments, prior evidence suggests that platforms remove only a limited fraction of hate content. In particular, disclosures by Frances Haugen in 2021 indicated that Facebook removed roughly 5\% of hate speech.

Because our empirical setting is Nigeria, we construct a benchmark by scaling global moderation costs by Nigeria’s share of platform users. With approximately 3 million Nigerian users among an estimated 350 million worldwide, Nigeria accounts for roughly 0.8\% of global users. Under the simplifying assumption that moderation resources scale proportionally with the number of users, this yields an implied Nigerian moderation budget of approximately USD 50,000 per month. While this allocation is unlikely to be exact, it provides a transparent baseline in the absence of country-level moderation data.

Our intervention cost approximately USD 7,500 per month and was associated with a reduction of roughly 2.5\% in hateful posting and reposting. Dividing monthly program costs by the estimated percentage reduction in hate suggests that the intervention achieved reductions at approximately 71\% lower cost per percentage point than conventional moderation. This comparison abstracts from important differences in mechanisms, ex-post removal versus ex-ante behavior change, but suggests that targeted, prosocial messaging may represent a comparatively cost-effective complement to conventional moderation.

%\section{Deviations from pre-registration}
\end{appendices}
\onehalfspacing
\end{document}